\documentclass[prd,aps,nofootinbib,notitlepage,showpacs,preprintnumbers,twocolumn]{revtex4-1}
\usepackage{graphicx,epsf,amsmath,amsfonts,amssymb,amsbsy}
\usepackage{epsfig}%\usepackage[cp1251]{inputenc}
\usepackage[T2A]{fontenc}
\usepackage{epstopdf}
\newcommand{\ds}{\displaystyle}
\newcommand{\vev}[1]{\langle#1\rangle}
\newcommand{\mat}{\left ( \begin{array}}
\newcommand{\emat}{\end{array} \right )}
\newcommand{\vect}{\left ( \begin{array}{c}}
\newcommand{\evect}{\end{array} \right )}

\newcommand{\Det}{\mathop{\rm Det}\nolimits}
\newcommand{\e}{\mathop{\rm e}\nolimits}
\begin{document}

%\hfill HU-EP-11/11
\title{ \bf \Large
Inhomogeneous charged pion
condensation in chiral asymmetric
dense quark matter in the framework of NJL$_2$ model  }

\author{
T. G. Khunjua $^{1)}$, K. G. Klimenko $^{2)}$, R. N. Zhokhov $^{2)}$, and V. C. Zhukovsky $^{1)}$}
\vspace{1cm}

\affiliation{$^{1)}$ Faculty of Physics, Moscow State University,
119991, Moscow, Russia} \affiliation{$^{2)}$ State Research Center
of Russian Federation -- Institute for High Energy Physics,
NRC "Kurchatov Institute", 142281, Protvino, Moscow Region, Russia}

\begin{abstract}
In this paper we investigate the phase structure of a
(1+1)-dimensional quark model with four-quark interaction
and in the presence of baryon ($\mu_B$), isospin ($\mu_I$) and chiral isospin ($\mu_{I5}$) chemical potentials. Spatially inhomogeneous chiral density wave (for chiral condensate) and single wave (for charged pion condensate) approaches are used. It is established that in the large-$N_c$ limit ($N_c$ is the number of colored quarks) there exists a duality correspondence
between the chiral symmetry breaking phase and the charged pion
condensation (PC) one. Moreover, it is shown that inhomogeneous
charged PC phase with nonzero baryon density is induced in the model
by arbitrary small values of the chemical potential $\mu_{I5}$ (for a
rather large region of $\mu_B$ and $\mu_I$).
\end{abstract}
%\pacs{12.39.Ki, 12.38.Mh, 21.65.Qr}

%%% 12.38.Mh Quark-gluon plasma
%%% 21.65.Qr Quark matter
%%% 12.39.Ki Relativistic quark model

\maketitle
\section{Introduction}
%$\dd$

QCD at nonzero temperature and baryon chemical potential plays a fundamental role in
the description of a number of various physical systems. Two important ones are neutron stars, which probe the low temperature and intermediate baryon chemical
potential domain, and heavy ion collision experiments, which explore
the region of the high temperature and low baryon chemical potential domain. However, the consideration of these systems is not possible in the framework of perturbative weak coupling QCD. Calculations with nonzero baryonic chemical potential $\mu_B$ is very hard to be performed on the lattice as well. Standard Monte-Carlo simulations are only possible for zero or
small values of $\mu_B$ because an evaluation of the QCD partition
function requires taking a path integral with a measure which includes
a complex fermion determinant (it is called sign problem). These are
the main reasons why our understanding of QCD at finite baryon density
is still rudimentary. Many interesting phenomena, such as
color superconductivity and color-flavor locking, etc, might occur at finite
baryon density, i.e. beyond the reach of current lattice and
perturbative QCD techniques.

%%%%%%%%%,
To describe physical situations, when the baryonic density is nonzero
but is comparatively low, usually different effective theories are employed.
Among them, we especially would like to mention the NJL-type models \cite{njl}. In this way, QCD phase diagrams
including chiral symmetry restoration \cite{asakawa,ebert,sadooghi,boer}, color superconductivity \cite{alford,klim,incera}, and charged pion condensation (PC) phenomena \cite{son,eklim,ak,mu,andersen,mam,car,villa} were investigated under heavy-ion experimental and/or compact star conditions, i.e. in the presence of finite temperature $T$, different chemical potentials and possible external (chromo)magnetic fields. 
There are other low-energy effective theories for QCD alternative to NJL model. One of them is the quark-meson model, or linear sigma model with quarks, which shares many features with the NJL model, but is renormalizable. More details about the properties of the quark-meson model can be found, e.g., in the reviews \cite{scad,and} and recent papers \cite{and2}. 
Also worth mentioning is the NJL model extended by Polyakov loop. In contrast to the usual NJL model, it mimics the features of confinement by coupling a nontrivial background gauge field to quarks (see, e.g., the review \cite{and}). However, consideration of the QCD phase diagram in terms of the quark-meson and Polyakov-loop NJL models is beyond the scope of our paper. We restrict ourselves to discussing only the properties of NJL models. They are 
nonrenormalizable in (3+1)-dimensional spacetime and can be considered only as effective field theories. This means that in the framework of NJL$_4$ models one can describe only phenomena at {\it comparatively low} energies,
temperatures and densities (chemical potentials). 

But there exist also low-dimensional theories, such as (1+1)-dimensional chiral Gross--Neveu (GN) type models \cite{gn,ft}, 
\footnote{Below we shall use the notation ``NJL$_2$ model'' instead of ``chiral GN model'' for (1+1)-dimensional models with {\it continuous chiral and/or isotopic, etc, symmetries}, since the chiral structure of the Lagrangian is the same as that of the corresponding (3+1)-dimensional NJL model.} that possess a lot of common features with QCD. For example, renormalizability, asymptotic freedom, dimensional transmutation, the spontaneous breaking of chiral symmetry (in vacuum) are the properties of the QCD and NJL$_2$ models \cite{wolff,kgk1,barducci,chodos}. 
In addition, they have the similar $\mu_B-T$ phase diagrams.
Hence, NJL$_2$ type models can be used as a laboratory for the qualitative simulation of specific properties of QCD at {\it arbitrary energies}. It is currently well understood (see, e.g., the discussion in \cite{barducci,chodos,thies}) that the usual {\it no-go} theorem \cite{coleman}, which generally forbids the spontaneous
breaking of any continuous symmetry in two-dimensional spacetime, does
not work in the limit  $N_c\to\infty$, where $N_c$ is the number of
colored quarks. This follows directly from the fact that in the limit
of large $N_c$ the quantum fluctuations, which would otherwise destroy
a long-range order corresponding to a spontaneous symmetry breaking,
are suppressed by $1/N_c$ factors. Thus,  the effects inherent for
real dense quark matter, such as chiral symmetry breaking phenomenon
(spontaneous breaking of the continuous axial $U(1)$ symmetry) or
charged pion condensation (spontaneous breaking of the continuous
isospin symmetry) might be simulated in terms of a simpler
(1+1)-dimensional NJL-type model, though only in the leading order of
the large $N_c$ approximation (see, e.g., Refs. \cite{thies} and
\cite{ektz,massive,ek,gubina,gkkz,andersen2}, respectively).

Besides the temperature and the baryon density, there are additional
parameters, which may be relevant for the above mentioned QCD systems.
Such an important parameter is, for instance, an isotopic chemical
potential $\mu_I$. It allows to consider systems with isospin
imbalance (different numbers of $u$ and $d$ quarks). It is realized, e.g.,
in neutron stars, heavy-ion experiments, etc. So QCD phase diagram in
the presence of both baryonic and isotopic chemical potentials has
been recently a subject of intensive research in the framework of some
effective theories \cite{son,eklim,ak}, where the possibility of the
charged PC phase just at $\mu_I\ne 0$ was predicted.
However, the existence of the charged PC phase is established there
without sufficient certainty. Indeed, for some values of model
parameters (the coupling constant $G$, cutoff parameter $\Lambda$, etc.)
the charged PC phase with {\it nonzero baryon density} is allowed by
NJL$_4$ models. However, it is forbidden in the framework of the NJL$_4$ models for other physically interesting values of $G$ and $\Lambda$
\cite{eklim}. Moreover, if the electric charge neutrality constraint
is imposed, the charged pion condensation phenomenon  depends strongly
on the bare (current) quark mass values. In particular, it turns out
that the charged PC phase with {\it nonzero baryonic density} is
%%%%%%%%%%%if
forbidden in the framework of NJL$_4$ models if the bare quark masses
%%%%%%%%%%%%%%%%%
reach the physically acceptable values of $5\div 10$ MeV (see in
Ref. \cite{andersen}). Due to these circumstances, the question arises,
whether there exist any factors promoting the appearance of charged PC
phenomenon in dense baryonic matter.

A positive answer to this question was obtained in papers
\cite{ekkz,gkkz,ekk}. Indeed, it was shown in Refs. \cite{ekkz,gkkz}
%%%%%%%%%%%%%a
that a charged PC phase might be realized in a dense baryonic system
%%%%%%%%%%%%%%%%%%%5
with finite size or in the case of a spatially inhomogeneous
condensate of charged pions. These conclusions are demonstrated in
\cite{ekkz,gkkz} in the large-$N_c$ limit, using a (1+1)-dimensional
toy model with four-quark interactions and containing baryon and
isospin chemical potentials. Moreover, it was shown in \cite{ekk} in
the framework of the same toy NJL$_2$ model that this phase can be
realized if we take into account a nonzero chiral isotopic potential
in addition. This means that there should be chiral imbalance in the
system. Recall that chiral imbalance, i.e. a nonzero difference
between densities of left- and right-handed fermions, may arise from
the chiral anomaly in the quark-gluon-plasma phase of QCD and possibly
leads to the chiral magnetic effect \cite{fukus} in heavy-ion
collisions. It might be realized also in compact stars or condensed
matter systems \cite{andrianov} (see also the review \cite{ms}). Note
also that phenomena, connected with a chiral imbalance, are usually
described in the framework of NJL models with a chiral chemical
potential \cite{andrianov}. It was also shown in \cite{ekk} that in
order to realize charged PC phase in dense quark matter, this chiral
asymmetry should be rather large. However in  \cite{ekk} only the case of
homogeneous condensates was considered.

In contrast, in this paper we study the phase structure of the same (1+1)-dimensional NJL model with an additional assumption of the presence of spatially
inhomogeneous condensates. For simplicity, we take into account the
condensate inhomogeneity in the form of the chiral density wave for
chiral condensate and the single plane wave for charged pion
condensate.

The existence of spatially inhomogeneous phases in dense systems is certainly not a new idea. In condensed matter physics charge and spin density waves are
commonly found  (for a review see, e.g., \cite{grun}), and inhomogeneous crystalline
phases have also been discussed long time ago for superconductors by
Fulde and Ferrell, as well as by Larkin and Ovchinnikov \cite{ferrel,larkin} (this phase is often called LOFF phase). More recently the crystalline phase for color superconductors was considered in \cite{araj} (see also reviews \cite{cassalb,anglani}). Deryagin, Grigoriev, and Rubakov have shown that at high densities in
%%%%%%%%%%%%%%%%%%%%%%%%%%%%%%%%%%%%an ,
 the limit of an infinite number of colors $N_{c}$ the QCD ground state might be inhomogeneous and anisotropic so that the ground state has the
%%%%%%%%%%%%%%%%%%%%%%%%%
structure of the standing wave \cite{dgr}. It is very challenging to
find inhomogeneous condensate as a solution and find its form analytically. However, more often one just assume some ansatz with several parameters and then solve a minimax problem with respect to these parameters. With this idea in mind, the simplest possible ansatz is given by a single plane wave in some sense. In analogy with the spin-density waves in condensed matter systems \cite{overhauser1},
this ansatz is called ``chiral density wave`` (CDW) or ``dual chiral
density wave``, where ``dual`` refers to the presence of two (scalar
and pseudoscalar) standing waves \cite{nakano}. The ansatz is also
sometimes called ``chiral spiral`` because it describes a spiral. The
last term is often used for the (1+1)-dimensional case, for which it
was originally introduced in \cite{thies}, but we will rather use CDW. Being an analytically treatable case, CDW ansatz has been the object of intense
investigations during the course of the last 25 years and provides us
with an excellent prototype for many generic features of inhomogeneous
%%%%%%%%%%%%%%%%%%a
condensation in quark matter. There can be a more favorable form of
%%%%%%%%%%%%%%%%%%%%%
condensate that minimizes thermodynamic potential even more
effectively, but investigating CDW we can at least conclude that
system favours inhomogeneity in some regions. Then one can try another
%%%%%%%%%%%%%%%%a
ansatz and find a deeper vacuum. Sometimes one can find many
%%%%%%%%%%%%%%%%%%%%%%%
inhomogeneous condensates and then find the most favorable. In some
models the phase structure with inhomogeneous condensates can be very
rich. The modern state of investigations of dense baryonic matter in
the framework of inhomogeneous condensate approach is presented in the
recent review \cite{buballa,Heinz:2014pkl} (see also, e.g., the recent
papers \cite{zfk,heinz,heinz2}).

In this paper we investigate the possibility of formation of inhomogeneous
condensates in the system and its influence on the phase diagram and
charged PC phenomenon in the framework of an extended (1+1)-dimensional NJL model with two quark flavors and in the presence of the baryon ($\mu_B$), isospin ($\mu_I$)  as well as chiral isospin ($\mu_{I5}$) chemical potentials. Note that earlier the phase structure of this model both in the framework of homogeneous approach for condensates and with inhomogeneous one (using CDW ansatz for quark
condensate and LOFF single plane wave ansatz for charged pion condensate) was investigated in \cite{gubina,massive,ek,ektz,gkkz,andersen2} at $\mu_{I5}=0$, i.e. at zero chiral asymmetry of quark matter. We now consider the extension of this model to the case of $\mu_{I5}\ne 0$. We will see that inhomogeneity is quite favored and is realized in almost the whole range of parameters  in the considered model. We will show that inhomogeneity of condensates does not change the fact that a chiral imbalance of dense and isotopically asymmetric baryon matter is a factor, which can induce there a charged
%%%%%%%%%%%%%%%thethe,
 PC phase. Moreover, this generation in the inhomogeneous case is even enhanced. It will be shown that in the inhomogeneous case charged PC phase is realized, in
%%%%%%%%%%%%%%%%%%%
 comparison with the results of the paper \cite{ekk}, even at small chiral asymmetry.

Moreover, it has been shown in the framework of the NJL$_2$ model
under consideration that in the leading order of the large-$N_c$
approximation there arises a duality between chiral symmetry breaking
(CSB) and charged PC phenomena. It means that if at $\mu_I=A$ and
$\mu_{I5}=B$ (at arbitrary fixed chemical potential $\mu_B$), e.g.,
the CSB (or the charged PC) phase is realized in the model, then at
the permuted values of  these chemical potentials, i.e. at $\mu_I=B$
and $\mu_{I5}=A$, the charged PC (or the CSB) phase is arranged. So,
it is enough to know the phase structure of the model at
$\mu_I<\mu_{I5}$, in order to establish the phase structure at
$\mu_I>\mu_{I5}$. Knowing condensates and other dynamical and
thermodynamical quantities of the system, e.g. in the CSB phase, one
can then obtain the corresponding quantities in the dually conjugated
charged PC phase of the model, by simply performing there the duality
transformation, $\mu_I\leftrightarrow\mu_{I5}$. \footnote{Note that
another kind of duality correspondence, the duality between CSB and
superconductivity, was demonstrated both in (1+1)- and
(2+1)-dimensional NJL models \cite{thies2,ekkz2}.} This feature of
the model does not depend on whether condensate is homogeneous or
inhomogeneous. This duality was noted in the paper \cite{ekk}, where
homogeneous condensates were considered.

The paper is organized as follows. In Sec. II a toy (1+1)-dimensional NJL-type model
with two quark flavors ($u$ and $d$ quarks) and including three kinds
of chemical potentials, $\mu_B,\mu_I,\mu_{I5}$, is presented. Next,
the symmetries of the model are discussed and the unrenormalized
thermodynamic potential (TDP) of the model under consideration is
obtained in the leading order of the large-$N_c$ expansion in the case
of inhomogeneous condensates. Here the dual symmetry of the model TDP
is established. It means that it is invariant under the simultaneous
%%%%%%%%%%%%%%%%%%%%%%%%%%%,
interchange of $\mu_I,\mu_{I5}$ chemical potentials%, and 
as well as chiral and charged pion condensates. In Sec. III  the renormalization of the TDP
is performed in the case of homogeneous ansatz for condensates. In
Sec. IV inhomogeneous case is considered. In Sec. IV A it is explained
how to obtain thermodynamic potential for inhomogeneous case from the
one for homogeneous case and it is argued that in order to get
physical thermodynamic potential, the subtraction procedure has to be applied. In Sec IV B different phase portraits of the model are obtained. Moreover, here the role of duality between chiral symmetry breaking and charged pion condensation phenomenon and its influence on the phase diagram are established. Sec VI contains summary and conclusions. Some technical
details are relegated to Appendix A.

\section{ The model and its thermodynamic potential}

We consider a two-dimensional model which is intended for simulation of the
properties of real dense quark matter with two massless quark flavors
($u$ and $d$ quarks). Its Lagrangian, which is symmetrical under
global color $SU(N_c)$ group, has the form
\begin{align}%{eqnarray}
& L=\bar q\Big [\gamma^\nu\mathrm{i}\partial_\nu
+\frac{\mu_B}{3}\gamma^0+\frac{\mu_I}2 \tau_3\gamma^0+\frac{\mu_{I5}}2
\tau_3\gamma^0\gamma^5\Big ]q\nonumber\\
&+\frac{G}{N_c}\Big [(\bar qq)^2+(\bar q\mathrm{i}\gamma^5\vec\tau q)^2 \Big
],  \label{1}
\end{align}%{eqnarray}
where the quark field $q(x)\equiv q_{i\alpha}(x)$ is a flavor doublet ($i=1,2$ or $i=u,d$) and color $N_c$-plet ($\alpha=1,...,N_c$) as well as a two-component Dirac spinor (the summation in (\ref{1}) over flavor, color, and spinor indices is implied); $\tau_k$ ($k=1,2,3$) are Pauli matrices. The quantities $\gamma^\nu$ ($\nu =0,1$) and $\gamma^5$ in Eq. (1) are matrices in the two-dimensional spinor space,
\begin{equation}
\begin{split}
\gamma^0=\begin{pmatrix}
0&1\\
1&0\\
\end{pmatrix};%\quad
\gamma^1=\begin{pmatrix}
0&-1\\
1&0\\
\end{pmatrix};%\quad
\gamma^5=\gamma^0\gamma^1=
\begin{pmatrix}
1&0\\
0&{-1}\\
\end{pmatrix}.
\end{split}\label{02}
\end{equation}
It is evident that the model (\ref{1}) is a generalization of the (1+1)-dimensional Gross-Neveu model \cite{gn} with a single massless quark color $N_c$-plet to the case of two quark flavors and additional baryon $\mu_B$, isospin $\mu_I$ and axial isospin $\mu_{I5}$ chemical potentials. These parameters are introduced in order to describe in the framework of the model (1) quark matter with nonzero baryon $n_B$, isospin $n_I$ and axial isospin $n_{I5}$ densities, respectively. The quantities $n_B$, $n_I$ and $n_{I5}$ are densities of conserved charges, which correspond to the invariance of Lagrangian (1) with respect to the abelian $U_B(1)$, $U_{I_3}(1)$ and $U_{AI_3}(1)$ groups, where \footnote{\label{f1,1}
Recall for the following that~~
$\exp (\mathrm{i}\alpha\tau_3)=\cos\alpha
+\mathrm{i}\tau_3\sin\alpha$,~~~~
$\exp (\mathrm{i}\alpha\gamma^5\tau_3)=\cos\alpha
+\mathrm{i}\gamma^5\tau_3\sin\alpha$.}
\begin{align}%{eqnarray}
&U_B(1):~q\to\exp (\mathrm{i}\alpha/3) q;~\nonumber\\
&U_{I_3}(1):~q\to\exp (\mathrm{i}\alpha\tau_3/2) q;~\nonumber\\
&U_{AI_3}(1):~q\to\exp (\mathrm{i}
\alpha\gamma^5\tau_3/2) q.
\label{2001}
\end{align}%{eqnarray}
So we have from Eq. (\ref{2001}) that $n_B=\bar q\gamma^0q/3$, $n_I=\bar q\gamma^0\tau^3 q/2$ and $n_{I5}=\bar q\gamma^0\gamma^5\tau^3 q/2$. We would like also to remark that, in addition to Eq. (\ref{2001}), Lagrangian (1) is invariant with respect to the electromagnetic $U_Q(1)$ group,
\begin{eqnarray}
U_Q(1):~q\to\exp (\mathrm{i}Q\alpha) q,
\label{2002}
\end{eqnarray}
where $Q={\rm diag}(2/3,-1/3)$ is the matrix of electric charges of
$u$ and $d$ quarks. The goal of the present paper is investigating the properties
of the ground state (the state of thermodynamic equilibrium)
of the system (1), i.e. the phase structure of the model and its dependence on the chemical potentials  $\mu_B$, $\mu_I$ and $\mu_{I5}$. (Note that at $\mu_{I5}=0$ the phase structure of this model was already investigated in details, e.g., in Refs \cite{gubina,massive,ek,ektz,gkkz,andersen2}.) So, we should (i) find the TDP of the system, (ii) determine its global minimum point (GMP), and (iii) investigate  the GMP dependence vs chemical potentials $\mu_B$, $\mu_I$ and $\mu_{I5}$. The ground state expectation values of $n_B$,  $n_I$ and $n_{I5}$ can be found by differentiating the TDP of the system (1) with respect to the corresponding chemical potential.

To find the thermodynamic potential of the system, we use a semi-bosonized version of the Lagrangian (\ref{1}), which contains composite bosonic fields $\sigma (x)$ and $\pi_a (x)$ $(a=1,2,3)$ (in what follows, we use the notations $\mu\equiv\mu_B/3$, $\nu=\mu_I/2$ and $\nu_{5}=\mu_{I5}/2$):
\begin{align}%{eqnarray}
&\widetilde L\ds =\bar q\Big [\gamma^\rho\mathrm{i}\partial_\rho
+\mu\gamma^0+ \nu\tau_3\gamma^0+\nu_{5}\tau_3\gamma^1-\sigma
-\mathrm{i}\gamma^5\pi_a\tau_a\Big ]q\nonumber\\
 &~~~-\frac{N_c}{4G}\Big [\sigma\sigma+\pi_a\pi_a\Big ].
\label{2}
\end{align}%{eqnarray}
In Eq. (\ref{2}) the summation over repeated indices is implied. In addition, we take into account there the relation $\gamma^0\gamma^5=\gamma^1$, which follows from Eq. (\ref{02}). From the Lagrangian (\ref{2}) one gets the equations for the bosonic fields
\begin{eqnarray}
\sigma(x)=-2\frac G{N_c}(\bar qq);~~~\pi_a (x)=-2\frac G{N_c}(\bar q
\mathrm{i}\gamma^5\tau_a q).
\label{200}
\end{eqnarray}
Note that the composite bosonic field $\pi_3 (x)$ can be identified with the physical $\pi_0$ meson, whereas the physical $\pi^\pm (x)$-meson fields are the following combinations of the composite fields, $\pi^\pm (x)=(\pi_1 (x)\pm i\pi_2 (x))/\sqrt{2}$. Obviously, the semi-bosonized Lagrangian $\widetilde L$ is equivalent to the initial Lagrangian (\ref{1}) when using the equations (\ref{200}). Furthermore, it is clear from (\ref{2001}), (\ref{200}) and footnote \ref{f1,1} that the bosonic fields transform under the isospin $U_{I_3}(1)$ and axial isospin $U_{AI_3}(1)$ groups in the following manner:
\begin{align}%{eqnarray}
&U_{I_3}(1):~\sigma\to\sigma;~~\pi_3\to\pi_3;~~\nonumber\\
&\pi_1\to\cos
(\alpha)\pi_1+\sin (\alpha)\pi_2;~~\pi_2\to\cos (\alpha)\pi_2-\sin
(\alpha)\pi_1.\nonumber\\
&U_{AI_3}(1):~\pi_1\to\pi_1;~~\pi_2\to\pi_2;~~\sigma\to\cos
(\alpha)\sigma+\sin (\alpha)\pi_3;~~\nonumber\\
&\pi_3\to\cos
(\alpha)\pi_3-\sin (\alpha)\sigma.%~~~~~~~~~~~201
\label{201}
\end{align}%{eqnarray}
Starting from the theory (\ref{2}), one obtains in the leading order of the large $N_c$-expansion (i.e. in the one-fermion loop approximation) the following path integral expression for the effective action ${\cal S}_{\rm {eff}}(\sigma,\pi_a)$ of the bosonic $\sigma (x)$ and $\pi_a (x)$ fields:
$$
\exp(\mathrm{i}{\cal S}_{\rm {eff}}(\sigma,\pi_a))=
  N'\int[d\bar q][dq]\exp\Bigl(\mathrm{i}\int\widetilde L\,d^2 x\Bigr),
$$
where
\begin{equation}
{\cal S}_{\rm {eff}}
(\sigma,\pi_a)
=-N_c\int d^2x\left [\frac{\sigma^2+\pi^2_a}{4G}
\right ]+\tilde {\cal S}_{\rm {eff}}
\label{3}
\end{equation}
and $N'$ is a normalization constant. The quark contribution to the effective action, i.e. the term $\tilde {\cal S}_{\rm {eff}}$ in Eq. (\ref{3}), is given by:
\begin{align}%{equation}
&\exp(\mathrm{i}\tilde {\cal S}_{\rm {eff}})=N'\int [d\bar
q][dq]\exp\Bigl(\mathrm{i}\int\Big\{\bar q\big
[\gamma^\rho\mathrm{i}\partial_\rho +\mu\gamma^0\nonumber\\
&+\nu\tau_3\gamma^0+
\nu_5\tau_3\gamma^1-\sigma -\mathrm{i}\gamma^5\pi_a\tau_a\big
]q\Big\}d^2 x\Bigr).
 \label{4}
\end{align}%{equation}
The ground state expectation values  $\vev{\sigma(x)}$ and $\vev{\pi_a(x)}$ of the composite bosonic fields are determined by
the saddle point equations,
\begin{eqnarray}
\frac{\delta {\cal S}_{\rm {eff}}}{\delta\sigma (x)}=0,~~~~~
\frac{\delta {\cal S}_{\rm {eff}}}{\delta\pi_a (x)}=0,~~~~~
\label{05}
\end{eqnarray}
where $a=1,2,3$. It is clear from Eq. (\ref{201}) that if $\vev{\sigma(x)}\ne 0$ and/or $\vev{\pi_3(x)}\ne 0$, then the axial isospin $U_{AI_3}(1)$ symmetry of the model is spontaneously broken down, whereas at $\vev{\pi_1(x)}\ne 0$ and/or $\vev{\pi_2(x)}\ne 0$ we have a spontaneous breaking of the isospin $U_{I_3}(1)$ symmetry. Since in the last case the ground state expectation values, or condensates, both of the field $\pi^+(x)$ and of the field $\pi^-(x)$ are nonzero, this phase is usually called the charged pion condensation (PC) phase. It is easy to see from Eq. (\ref{200}) that the nonzero condensates $\vev{\pi_{1,2}(x)}$ (or $\vev{\pi^\pm(x)}$) are not invariant with respect to the electromagnetic $U_Q(1)$ transformations (\ref{2002}) of the flavor quark doublet. Hence in the charged PC phase the electromagnetic $U_Q(1)$ invariance of the model (1) is also broken  spontaneously, so superconductivity is an unavoidable property of the charged PC phase.

In vacuum, i.e. in the state corresponding to an empty space with zero particle density and zero values of the chemical potentials $\mu$, $\nu$ and $\nu_5$, the quantities $\vev{\sigma(x)}$ and $\vev{\pi_a(x)}$ do not depend on space coordinate $x$. However, in a dense medium, when $\mu\ne 0$, $\nu\ne 0$ and $\nu_5\ne 0$, the ground state expectation values of bosonic fields might have a nontrivial dependence on the spatial coordinate $x$. In particular, in this paper we use the following spatially inhomogeneous CDW ansatz for chiral condensate and the single plane wave ansatz for charged pion condensates:
\begin{align}%{eqnarray}
&\vev{\sigma(x)}=M\cos (2kx),~~~\vev{\pi_3(x)}=M\sin
(2kx),~~~\nonumber\\
&\vev{\pi_1(x)}=\Delta\cos(2k'x),~~~
\vev{\pi_2(x)}=\Delta\sin(2k'x), \label{06}
%~~~~~~~~~06
\end{align}%{eqnarray}
where gaps $M,\Delta$ and wavevectors $k,k'$ are constant dynamical quantities. In fact, they are coordinates of the global minimum point (GMP) of the thermodynamic potential (TDP) $\Omega (M,k,k',\Delta)$.
\footnote{Here and in what follows we will use a rather conventional notation "global" minimum in the sense that among all our numerically found local minima the TDP takes in their case the lowest value. This does not exclude the possibility that there exist other inhomogeneous condensates, different from (\ref{06}), which lead to ground states with even lower values of the TDP.}
In the leading order of the large $N_c$-expansion it is defined by the following expression:
\begin{align}%{equation}
&\int d^2x \Omega (M,k,k',\Delta)\nonumber\\
&=-\frac{1}{N_c}{\cal S}_{\rm {eff}}\{\sigma(x),\pi_a(x)\}\big|_{\sigma
    (x)=\vev{\sigma(x)},\pi_a(x)=\vev{\pi_a(x)}} ,
\end{align}%{equation}
which gives
\begin{align}%{equation}
&\int d^2x\Omega (M,k,k',\Delta)=\int
d^2x\frac{M^2+\Delta^2}{4G}\nonumber\\&+\frac{\mathrm{i}}{N_c}\ln\left (
\int [d\bar q][dq]\exp\Bigl(\mathrm{i}\int d^2
x\bar q \widetilde{D} q \Bigr)\right ),
\label{08}
\end{align}%{equation}
where
\begin{align}%{equation}
&\bar q  \widetilde{D} q=\bar q\big
[\gamma^\rho\mathrm{i}\partial_\rho
+\mu\gamma^0+\nu\tau_3\gamma^0+\nu_5\tau_3\gamma^1\nonumber\\
&-M\exp(2\mathrm{i}\gamma^5\tau_3kx)\big ]q-\Delta\big (\bar q_u\mathrm{i}\gamma^5 q_d\big )\e^{-2\mathrm{i}k'x}\nonumber\\
&-\Delta\big (\bar q_d\mathrm{i}\gamma^5 q_u\big )\e^{2\mathrm{i}k'x}.\label{09}
\end{align}%{equation}{equation}
(Remember that in this formula $q$ is indeed a flavor doublet, i.e. $q=(q_u,q_d)^T$.) To proceed, let us introduce in Eqs (\ref{08})-(\ref{09}) the new quark doublets, $\psi$ and $\bar\psi$, by the so-called Weinberg (or chiral) transformation of these fields \cite{weinberg}, $\psi=\exp(\mathrm{i}\tau_3k'x+\mathrm{i}\tau_3\gamma^5kx)q$ and $\bar\psi = \bar q\exp(\mathrm{i}\tau_3\gamma^5kx-\mathrm{i}\tau_3k'x)$. Since this transformation of quark fields does not change the path integral measure in Eq. (\ref{08})  \footnote{Strictly speaking, performing Weinberg transformation of quark fields in Eq. (\ref{08}), one can obtain in the path integral measure a factor, which however does not depend on the dynamical variables $M$, $\Delta$, $k$, and $k'$. Hence, we ignore this unessential factor in the following calculations. Note that only in the case when there is an interaction between spinor and gauge fields there might appear a nontrivial, i.e. dependent on dynamical variables, path integral measure, generated by Weinberg transformation of spinors. This unobvious fact follows from the investigations by Fujikawa \cite{fujikawa}.}, the
expression (\ref{08}) for the TDP is easily transformed to the following one:
\begin{align}%{eqnarray}{eqnarray}
&\int d^2x\Omega (M,k,k',\Delta)=\int
d^2x\frac{M^2+\Delta^2}{4G}\nonumber\\
&+\frac{\mathrm{i}}{N_c}\ln\left (
\int [d\bar\psi][d\psi]\exp\Bigl(\mathrm{i}\int d^2
x\bar\psi D\psi \Bigr)\right ),
\label{010}\end{align}
where  instead of the $x-$dependent Dirac operator $\widetilde{D}$ a new $x-$independent operator $D$ appears
\begin{equation}
D=\gamma^\nu \mathrm{i}\partial_\nu -M +\mu\gamma^0+(\nu_5+k')\tau_3\gamma^1+(\nu+k)\tau_3\gamma^0-\mathrm{i}\Delta\tau_1\gamma^5.
\label{110}
\end{equation}
The expression (\ref{010}) for the TDP now takes the form
\begin{align}%{eqnarray}
&\Omega (M,k,k',\Delta)=\frac{M^2+\Delta^2}{4G}+\mathrm{i}\frac{{\rm
Tr}_{csfx}\ln D}{N_c\int d^2x}\nonumber\\
&=\frac{M^2+\Delta^2}{4G}+\mathrm{i}{\rm
Tr}_{sf}\int\frac{d^2p}{(2\pi)^2}\ln\overline{D}(p),
\label{11}
\end{align}%{eqnarray}
where $\overline{D}(p)=\not\!p +\mu\gamma^0 + \tilde\nu\tau_3\gamma^0+\tilde\nu_5\tau_3\gamma^1-M-\mathrm{i}\gamma^5\Delta\tau_1$,
$\tilde\nu=\nu+k$ and $\tilde\nu_5=\nu_5+k'$. The Tr-operation ${\rm
  Tr}_{csfx}$ in Eq. (\ref{11}) stands for the trace in color- ($c$),
spinor- ($s$), flavor- ($f$) as well as two-dimensional coordinate-
($x$) spaces, respectively, and ${\rm Tr}_{sf}$ is the respective
trace without color and $x-$spaces. The general relation
\begin{eqnarray}
{\rm Tr}_{sf}\ln\overline{D}(p)=\ln\Det\overline{D}(p)=\sum_i\ln\epsilon_i,\label{det}
\end{eqnarray}
where the summation over all four eigenvalues $\epsilon_i$ of the
 4$\times$4 matrix $\overline{D}(p)$
\begin{eqnarray}
\epsilon_{1,2,3,4}=-M\pm\sqrt{{\cal N}\pm 2\sqrt{\cal P}},
\label{8}
\end{eqnarray}
is implied.  Here
\begin{align}
&{\cal N}=(p_0+\mu)^2-p_1^2-\Delta^2+
\tilde\nu^2-\tilde\nu_{5}^2,\nonumber\\
&{\cal P}=\big [(p_0+\mu)\tilde\nu+p_1\tilde\nu_{5}\big ]^2-\Delta^2(\tilde\nu^2-\tilde\nu_{5}^2).
\end{align}
So we have from Eq. (\ref{11}) and Eq. (\ref{det})
\begin{eqnarray}
\hspace{-0.3cm}\Omega (M,k,k',\Delta)~&& =\frac{M^2+\Delta^2}{4G}+\mathrm{i}\int\frac{d^2p}{(2\pi)^2}\ln
P_4(p_0).
\label{9}
\end{eqnarray}
In Eq. (\ref{9}) we use the notations
\begin{eqnarray}
P_4(p_0)=\epsilon_1\epsilon_2\epsilon_3\epsilon_4=\eta^4-2a\eta^2-b\eta+c,
\label{91}%~~~~~~~~~~91
\end{eqnarray}
where $\eta=p_0+\mu$ and
\begin{eqnarray}
\hspace{-0.3cm}a&&=M^2+\Delta^2+p_1^2+\tilde\nu^2+\tilde\nu_{5}^2;~~b=8p_1\tilde\nu\tilde\nu_{5};\nonumber\\
\hspace{-0.3cm}c&&=a^2-4p_1^2(\tilde\nu^2+\tilde\nu_5^2)-4M^2\tilde\nu^2-4\Delta^2\tilde\nu_5^2-4\tilde\nu^2\tilde\nu_5^2.
\label{10}%~~~~~~~~~~~~~~10
\end{eqnarray}
It is evident from Eq. (\ref{10}) that the expression (\ref{9}) for the TDP is an even function over each of the variables $M$ and $\Delta$. In addition, it is invariant under each of the transformations $\mu\to-\mu$,  $\tilde\nu\to-\tilde\nu$, $\tilde\nu_5\to-\tilde\nu_5$. \footnote{Indeed, if simultaneously with $\mu\to-\mu$ we perform in the integral (\ref{9}) the $p_0\to-p_0$ and $p_1\to-p_1$ change of variables, then one can easily see that the expression (\ref{9}) remains intact. Finally, if only $\tilde\nu$ (only $\tilde\nu_5$) is replaced by $-\tilde\nu$ (is replaced by $-\tilde\nu_5$), we should transform $p_1\to-p_1$ in the integral (\ref{9}) in order to be convinced that the TDP remains unchanged. }
Hence, without loss of generality we can consider in the following only $\mu\ge 0$, $\tilde\nu\ge 0$, $\tilde\nu_5\ge 0$, $M\ge 0$, and $\Delta\ge 0$ values of these quantities. Moreover, the expression (\ref{9}) for the TDP is invariant with respect to the so-called duality transformation,
\begin{eqnarray}
{\cal D}:~~~~M\longleftrightarrow \Delta,~~\nu\longleftrightarrow\nu_5,~~k\longleftrightarrow k'.
 \label{16}%~~~~~~~~~~~16
\end{eqnarray}
It means that in the leading order of the large-$N_c$ approximation there is the so-called duality correspondence between chiral symmetry breaking (CSB) and charged PC phenomena (in details, see below in Sec. IV). Note that another kind of duality correspondence, the duality between CSB and superconductivity, was demonstrated both in (1+1)- and (2+1)-dimensional NJL models \cite{thies2,ekkz2}. In powers of $\Delta$ the fourth-degree polynomial $P_4(p_0)$ has the following form
\begin{align}%{eqnarray}
&P_4(p_0)\equiv\Delta^4-2\Delta^2(\eta^2-p_1^2-M^2+\tilde\nu_5^2-\tilde\nu^2)\nonumber\\
&~~~~~~~~~~+\big [M^2+(p_1-\tilde\nu_5)^2-(\eta+\tilde\nu)^2\big
]\nonumber\\
&~~~~~~~~~~\times\big [M^2+(p_1+\tilde\nu_5)^2-
(\eta-\tilde\nu)^2\big ]. \label{17}
\end{align}%{eqnarray}
Expanding the right-hand side of Eq. (\ref{17}) in powers of $M$, one can obtain an equivalent alternative expression for this polynomial. Namely,
\begin{align}%{eqnarray}
&P_4(p_0)\equiv M^4-2M^2(\eta^2-p_1^2-\Delta^2+\tilde\nu^2-\tilde\nu_5^2)\nonumber\\
&~~~~~~~~~~+\big [\Delta^2+(p_1-\tilde\nu)^2-(\eta+\tilde\nu_5)^2\big ]\nonumber\\
&~~~~~~~~~~\times\big
[\Delta^2+(p_1+\tilde\nu)^2-(\eta-\tilde\nu_5)^2\big ].\label{18}
\end{align}%d{eqnarray}
Note also that according to the general theorem of algebra, the polynomial $P_4(p_0)$ can be presented in the form
\begin{eqnarray}
\hspace{-0.4cm}P_4(p_0)\equiv (p_0-P_{01})(p_0-P_{02})(p_0-P_{03})(p_0-P_{04}), \label{170}%~~~~~~~~~~~~~~170
\end{eqnarray}
where $P_{01}$, $P_{02}$, $P_{03}$ and $P_{04}$ are the roots of this polynomial. This quantities are the energies of quasiparticle or quasiantiparticle excitations of the system. In particular, it follows from Eq. (\ref{17}) that at $\Delta=0$ the set of roots $P_{0i}$ looks like
\begin{align}%{eqnarray}
&\Big\{P_{01},P_{02},P_{03},P_{04}\Big\}\Big |_{\Delta=0}=\Big\{\pm\sqrt{M^2+(p_1-\tilde\nu_5)^2}\nonumber\\
&-\mu-\tilde\nu,
~~-\mu+\tilde\nu\pm\sqrt{M^2+(p_1+\tilde\nu_5)^2}\Big\}, \label{26}
\end{align}%{eqnarray}
whereas it is clear from Eq. (\ref{18}) that at $M=0$ it has the form
\begin{align}%{eqnarray}
&\Big\{P_{01},P_{02},P_{03},P_{04}\Big\}\Big
|_{M=0}=\Big\{\pm\sqrt{\Delta^2+(p_1-\tilde\nu)^2}\nonumber\\
&-\mu-\tilde\nu_5,~~
-\mu+\tilde\nu_5\pm\sqrt{\Delta^2+(p_1+\tilde\nu)^2}\Big\}. \label{27}
\end{align}%{eqnarray}
Taking into account in Eq. (\ref{9}) the relation (\ref{170}) as well as a rather general formula
\begin{eqnarray}
\int_{-\infty}^\infty dp_0\ln\big
(p_0-K)=\mathrm{i}\pi|K|,\label{int}
\end{eqnarray}
(obtained rigorously, e.g., in Appendix B of \cite{gkkz} and being true up to an infinite term independent of the real quantity $K$), it is possible to integrate there over $p_0$. So the {\it unrenormalized} TDP (\ref{9}) can be presented in the following form,
\begin{align}%{eqnarray}
&\Omega (M,k,k',\Delta)\equiv\Omega^{un} (M,k,k',\Delta)=
\frac{M^2+\Delta^2}{4G}\nonumber\\
&-\int_{-\infty}^\infty\frac{dp_1}{4\pi}\Big (|P_{01}|+|P_{02}|+|P_{03}|+|P_{04}|\Big ). \label{28}
\end{align}%{eqnarray}

\section{Homogeneous case of the ansatz (\ref{06}) for condensates: $k=0$, $k'=0$}

%%%%%%%%%%%%%%%%%%%%%%%%%%%%%%%%appears
The case $k=0$, $k'=0$ was investigated in details in \cite{ekk}, where we have shown that at finite nonzero values of $\mu_{I5}$ there might appear in the
%%%%%%%%%%%%
 model (1) a charged PC phase with nonzero baryon density (see Fig. 1). Since some approaches, expressions, etc from our paper \cite{ekk} are necessary when considering the inhomogeneous ansatz for condensates, in the present section we reproduce them  briefly.

\subsection{Thermodynamic potential in the vacuum case: $\mu=0,\nu=0,\nu_5=0$}

It is interesting first of all to find the TDP of the model (1) in vacuum, when  $k=0$, $k'=0$ and $\mu=0$, $\nu=0$, $\nu_5=0$. Since in this case the thermodynamic potential (\ref{28}) is usually called effective potential, we use for it the notation $V^{un} (M,\Delta)$. As a consequence of Eqs (\ref{9})-(\ref{10}), it is clear that at $\mu=\tilde\nu=\tilde\nu_5=0$ the effective potential $V^{un} (M,\Delta)$ looks like
\begin{align}%{eqnarray}
&V^{un} (M,\Delta)\nonumber\\
&=\frac{M^2+\Delta^2}{4G}
+2i\int\frac{d^2p}{(2\pi)^2}\ln\Big
[p_0^2-p_1^2-M^2-\Delta^2\Big ]\nonumber\\
&=\frac{M^2+\Delta^2}{4G}-\int_{-\infty}^\infty\frac{dp_1}{\pi}\sqrt{p_1^2+M^2+\Delta^2}. \label{25}%~~~~~~~~~~25
\end{align}%{eqnarray}
(To obtain the last expression in this formula, we have integrated there over $p_0$ according to the general relation (\ref{int}).) It is evident that the effective potential (\ref{25}) is an ultraviolet divergent
quantity. So, we need to renormalize it. This procedure consists of two steps: (i) First of all we need to regularize the divergent
%%%%%%%%%%%%%%
integral in Eq. (\ref{25}), i.e. we suppose there that $|p_1|<\Lambda$ and replace bare coupling constant $G$ by the new $\Lambda$-dependent coupling constant $G(\Lambda)$. (ii)  Second, we must suppose also that the coupling constant $G(\Lambda)$ depends on the cutoff parameter $\Lambda$ in such a way that in the limit $\Lambda\to\infty$ one obtains a finite expression for the effective potential.
%%%%%%%%%%%%%%

Following the step (i) of this procedure, we have
\begin{align}%{eqnarray}
&V^{reg} (M,\Delta)=\frac{M^2+\Delta^2}{4G(\Lambda)}-\frac 2\pi\int_{0}^\Lambda
dp_1\sqrt{p_1^2+M^2+\Delta^2},
\end{align}
which gives
\begin{align}
&V^{reg} (M,\Delta)=\frac{M^2+\Delta^2}{4G(\Lambda)}-\frac
1\pi\left\{\Lambda\sqrt{\Lambda^2+M^2+\Delta^2}\right.\nonumber\\
&\left.+(M^2+\Delta^2)\ln\frac{\Lambda+\sqrt{\Lambda^2
+M^2+\Delta^2}}{\sqrt{M^2+\Delta^2}}\right\}. \label{29}%~~~~~~~~~~29
\end{align}%{eqnarray}
Further, according to the step (ii) we suppose that in Eq. (\ref{29}) the  coupling constant $G(\Lambda)$ has the following $\Lambda$ dependence:
\begin{eqnarray}
\frac 1{4G(\Lambda)}=\frac 1\pi\ln\frac{2\Lambda}{m}, \label{30}
\end{eqnarray}
where $m$ is a new free massive parameter of the model, which appears instead of the dimensionless bare coupling constant $G$ (dimensional transmutation) and, evidently, does not depend on a normalization point, i.e. it is a renormalization invariant quantity. Substituting Eq. (\ref{30}) into Eq. (\ref{29}) and ignoring there an unessential term $(-\Lambda^2/\pi)$, we have in the limit $\Lambda\to\infty$ the finite and renormalization invariant expression for the effective potential,
\begin{eqnarray}
V_0 (M,\Delta)&=&\frac{M^2+\Delta^2}{2\pi}\left [\ln\left (\frac{M^2+\Delta^2}{m^2}\right )-1\right ]. \label{31}
\end{eqnarray}
%%%%%%%%%
It is evident that the parameter $m$ corresponds to the minimum value of $M$ of the effective potential (\ref{31}) with $\Delta =0$. Then everything is measured in units of this mass scale. \footnote{Formally, the effective potential (\ref{31}) in vacuum might have a minimum at $\Delta\ne 0$. However, this is not a physical situation because it was shown by Vafa and Witten in Ref. \cite{vafa} that global symmetries such as isospin and baryon number in vector-like gauge theories like QCD cannot be spontaneously broken in vacuum, i.e.at zero chemical potentials. Since the NJL model is particularly interesting as a low-energy effective theory for QCD, we do not consider the minima of (\ref{31}) with $\Delta\ne 0$. But at nonzero values of $\mu$, Vafa-Witten theorem is no longer applicable, so the phase with $\Delta\ne 0$, which is predicted by some NJL models, can be realized in dense quark matter. }
%%%%%%%%%%

\subsection{Renormalization of the TDP (\ref{28}) in the
general case: $\mu>0$, $\nu>0$, $\nu_5>0$}

To find a renormalized expression for the TDP (\ref{28}) at $k=0$ and $k'=0$ in the general case, i.e. at $\mu>0$, $\nu>0$ and $\nu_5>0$, we need first of all to regularize it. Here we use the so-called momentum space regularization,
\begin{align}%{eqnarray}
&\Omega^{reg} (M,\Delta)=
\frac{M^2+\Delta^2}{4G(\Lambda)}-
\int_{0}^\Lambda\frac{dp_1}{2\pi}\Big (\sum_{i=1}^4|p_{0i}|\Big)
\label{032}\\
%\Big (|p_{01}|+|p_{02}|+|p_{03}|+|p_{04}|\Big )\label{032}~~~~~~~032\\
&=\frac{M^2+\Delta^2}{4G(\Lambda)}-\int_{0}^\Lambda\frac{dp_1}{2\pi}\Big (\sum_{i=1}^4|p_{0i}|\Big )\Big |_{\mu=\nu=\nu_5=0}\nonumber\\
&-\int_{0}^\Lambda\frac{dp_1}{2\pi}\Big [\sum_{i=1}^4|p_{0i}|-\Big (\sum_{i=1}^4|p_{0i}|\Big )\Big |_{\mu=\nu=\nu_5=0}\Big ], \label{32}%~~~~~~~32
\end{align}%{eqnarray}
where the notation $p_{0i}$ is accepted for the quasiparticle energy $P_{0i}$ at $k=0$ and $k'=0$ ($i=1,...,4$). In addition, we took into account that the quantities $P_{0i}$ and $p_{0i}$ are even functions with respect to $p_1$ (see Appendix \ref{ApB}). In Appendix \ref{ApB} other properties of the quasiparticle energies $p_{0i}$, where $i=1,...,4$, are also presented. Since the asymptotic expansion (\ref{B9}) for the quantity $\sum_{i=1}^4|p_{0i}|$ does not depend on chemical potentials $\mu$, $\nu$ and $\nu_5$, it is clear that the second integral in Eq. (\ref{32}) converges in the limit  $\Lambda\to\infty$. Moreover, one can see that due to the relation (\ref{B10}) the first two terms on the right-hand side of Eq. (\ref{32}) are no more, than the regularized effective potential in vacuum (\ref{29}). So to obtain a finite expression for the unrenormalized TDP (\ref{28}), it is enough to use in Eq. (\ref{32}) the way of the previous subsection, where just these two terms, i.e. the vacuum effective potential, were renormalized by an appropriate behavior (\ref{30}) of the coupling constant $G(\Lambda)$. Taking the relation (\ref{30}) into account, we have in the limit $\Lambda\to\infty$ for the TDP $\Omega^{reg} (M,\Delta)$ the following expression
\begin{align}%{eqnarray}
&\Omega^{ren}
(M,\Delta)=V_0(M,\Delta)\nonumber\\
&-\int^\infty_{0}\frac{dp_1}{2\pi}\Big
(\sum_{i=1}^4|p_{0i}|
%\Big\{|p_{01}|+|p_{02}|+|p_{03}|+|p_{04}|
-4\sqrt{p_1^2+M^2+\Delta^2}\Big\},
\label{35}
\end{align}%{eqnarray}
where $V_0(M,\Delta)$ is the renormalized TDP (effective potential) (\ref{31}) of the model at $\mu=\nu=\mu_5=0$. Moreover, we have used in Eq. (\ref{35}) the relation (\ref{B10}) for the sum of quasiparticle energies in vacuum.

Let us denote by $(M_0,\Delta_0)$ the global minimum point (GMP) of the TDP (\ref{35}). Then, investigating the behavior of this point vs $\mu$, $\nu$ and $\nu_5$ it is possible to construct the $(\mu,\nu,\nu_5)$-phase portrait (diagram) of the model. A numerical algorithm for
finding the quasi(anti)particle energies  $p_{01}$, $p_{02}$, $p_{03}$,
and $p_{04}$ is elaborated in Appendix \ref{ApB}. Based on this, it can be shown numerically that GMP of the TDP can never be of the form $(M_0\ne 0,\Delta_0\ne 0)$. Hence, in order to establish the phase portrait of the model, it is enough to study the projections $F_1(M)\equiv\Omega^{ren} (M,\Delta=0)$ and $F_2(\Delta)\equiv\Omega^{ren}(M=0,\Delta)$ of the TDP (\ref{35}) to the $M$ and $\Delta$ axes, correspondingly. Taking into account the relations (\ref{26}) and (\ref{27}) for the quasiparticle energies $p_{0i}$  at $\Delta=0$ or $M=0$, it is possible  to obtain the following expressions for these quantities,
\begin{align}%{eqnarray}
&F_1(M)=\frac{M^2}{2\pi}\ln\left(\frac{M^2}{m^2}\right)
-\frac{M^2}{2\pi}-\frac{\nu_5^2}{\pi}-\theta (\mu+\nu-M)\frac{\cal A}{\pi}\nonumber\\
&-\theta
(|\mu-\nu|-M)\theta(\sqrt{(\mu-\nu)^2-M^2}-\nu_5)\frac{\cal B}{2\pi}\nonumber\\
&+\theta (\mu+\nu-M)\theta
  (\sqrt{(\mu+\nu)^2-M^2}-\nu_5)\frac{\cal C}{2\pi },\label{33}%~~~~~~~33
\end{align}%{eqnarray}
where
\begin{align}
&{\cal A}=(\mu+\nu)\sqrt{(\mu+\nu)^2-M^2}\nonumber\\
&-M^2\ln\frac{\mu+\nu+\sqrt{(\mu+\nu)^2-M^2}}{M},
\end{align}
\begin{align}
&{\cal B}=|\mu-\nu|\sqrt{(\mu-\nu)^2-M^2}
+\nu_5\sqrt{\nu_5^2+M^2}\nonumber\\
&-2|\mu-\nu|\nu_5-M^2\ln\frac{|\mu-\nu|+\sqrt{|\mu-\nu|^2-M^2}}{\nu_5+\sqrt{\nu_5^2+M^2}},
\end{align}
\begin{align}
&{\cal C}=(\mu+\nu)\sqrt{(\mu+\nu)^2-M^2}\nonumber\\
&+\nu_5\sqrt{\nu_5^2+M^2}-2(\mu+\nu)\nu_5\nonumber\\
&-M^2\ln\frac{\mu+\nu+\sqrt{(\mu+\nu)^2-M^2}}{\nu_5+\sqrt{\nu_5^2+M^2}}.
\end{align}%{eqnarray}
\begin{equation}%{eqnarray}
F_2(\Delta)=F_1(\Delta)\Bigg |_{\nu\longleftrightarrow\nu_5}. \label{34}
\end{equation}
(Details of the derivation of these expressions are given in Appendix
B of \cite{ekk}.) After simple transformations, it is clear that
$F_1(M)$ and $F_2(\Delta)$ coincide at $\nu_5=0$ with corresponding
TDPs (12) and (13) of the paper \cite{ek}.

To find the phase structure of the model (1), it is necessary to determine (numerically) the global minimum points of the TDPs $F_1(M)$ (\ref{33}) and $F_2(\Delta)$ (\ref{34}) and then compare the minimum values of these functions. The result is the GMP of the whole TDP (\ref{35}). Investigating the behavior of this GMP vs external parameters $\mu,\nu,\nu_5$, one can establish the phase structure of the model in the case of approach with spatially homogeneous condensates.

Moreover, the derivative of the TDP vs $\mu$ supplies the quark number density $n_q$ in each phase. In Fig. 1 the $(\mu,\nu,\nu_5)$-phase portrait of the model is presented in the supposition that all condensates are spatially homogeneous \cite{ekk}. It is clear from this figure that charged PC phase with nonzero quark number density $n_q$ (this phase is denoted there by PCd) can be realized in the model (1) only at rather large values of $\nu_5$.

Now we are ready to consider the role and influence of the chiral isotopic chemical potential $\nu_5$ on the phase structure of the model (1) in a more general approach, when condensates are spatially inhomogeneous and restricted by the ansatz (\ref{06}) with $k,k'\ne 0$.
\begin{figure*}
%----figure 1,2
\includegraphics[width=0.45\textwidth]{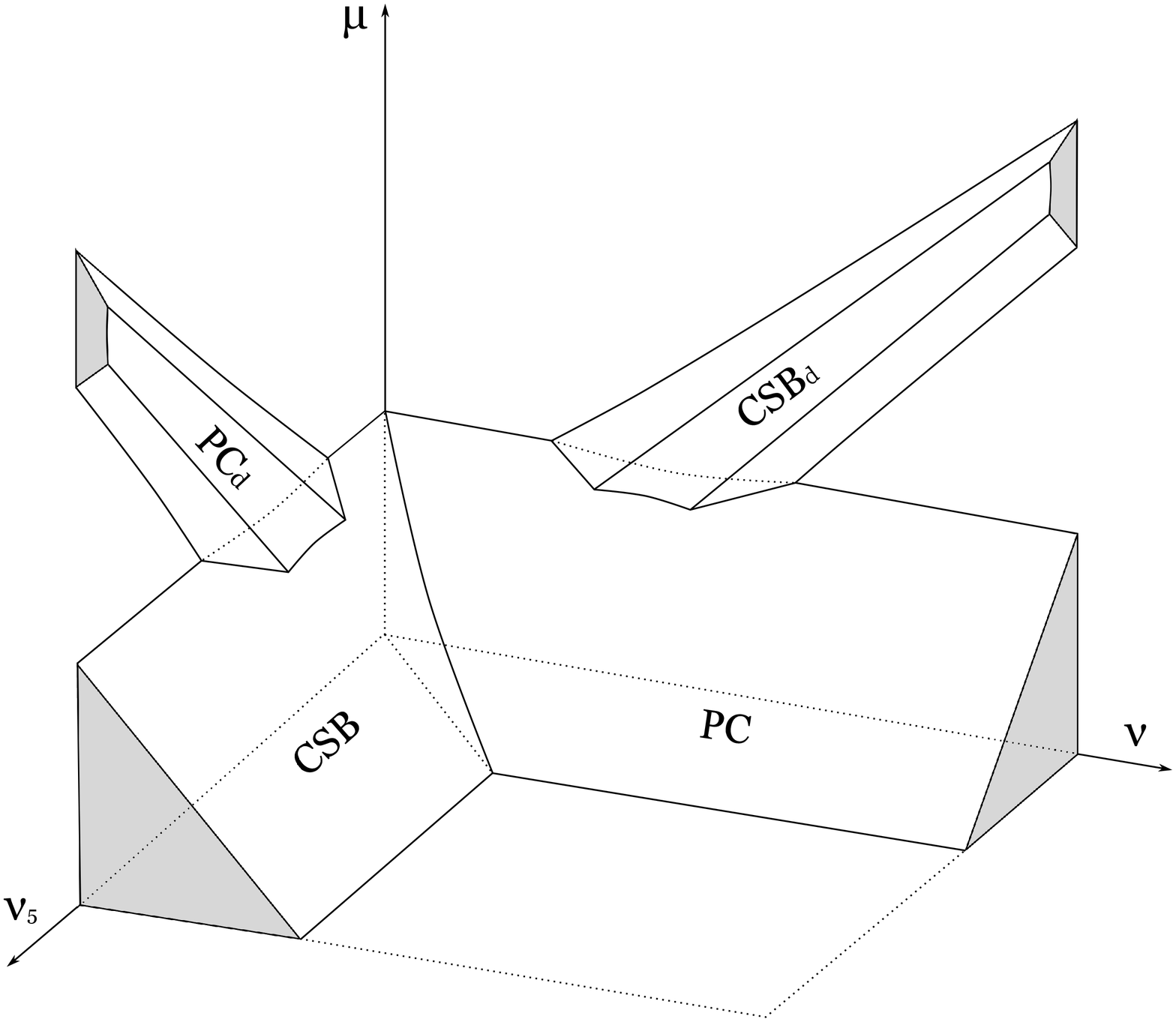}
\hfill
\includegraphics[width=0.45\textwidth]{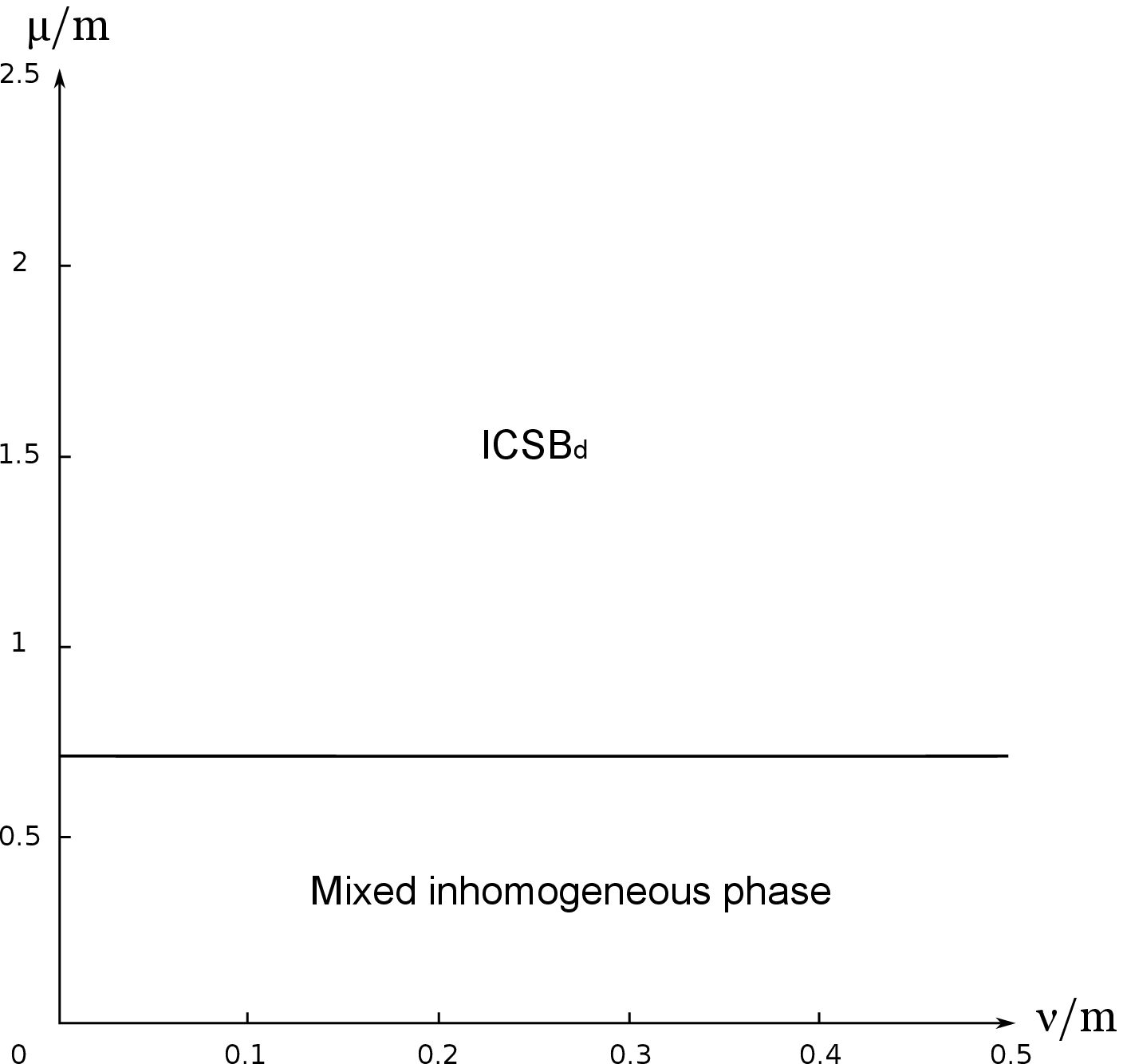}\\
\parbox[t]{0.45\textwidth}{\caption{Schematic representation of the $(\nu_5,\nu,\mu)$-phase portrait of the model in the case of spatially homogeneous condensates. It consists of charged pion condensation (PC) and chiral symmetry breaking (CSB) phases, in which quark number density is zero. Moreover, there are charged pion condensation (PCd) and chiral symmetry breaking (CSBd) phases with nonzero quark number density. The points which are outside PC-, CSB-, PCd-, and CSBd phases of the diagram correspond to the symmetric phase.}}\hfill
\parbox[t]{0.45\textwidth}{\caption{{\bf The case of spatially inhomogeneous condensates}: The $(\nu,\mu)$-phase portrait of the model at  $\nu_5=0_+$. The notation ICSBd means the inhomogeneous chiral symmetry breaking phase with nonzero quark number density.  For arbitrary point of the region ''Mixed inhomogeneous phase`` there is a degeneracy between global minima of the TDP corresponding to inhomogeneous chiral symmetry breaking and inhomogeneous charged pion condensation phases. Quark number density in both is zero. $m$ is a massive parameter introduced in Eq. (\ref{30}).}}
\end{figure*}

\section{Inhomogeneous case of the ansatz (\ref{06}): $k\ne 0$, $k'\ne 0$}

 \subsection{ Thermodynamic potential }

As it is clear from comments after the formula (\ref{11}), at $k,k'\ne 0$ the unrenormalized TDP (\ref{28}) $\Omega^{un}(M,k,k',\Delta)$ can be obtained from the unrenormalized TDP in the case of homogeneous condensates simply performing there the replacement $\nu, \nu_5\to\tilde{\nu}\equiv\nu+k, \tilde{\nu_5}\equiv\nu_5+k'$. Moreover, throughout the paper we suppose that  $\tilde{\nu}\ge 0, \tilde{\nu_5}\ge 0$ (see the motivation given after Eq. (23)). So, in order to obtain a finite (or renormalized) expression for the TDP at $k,k'\ne 0$, it is natural to use the same renormalized procedure as for the case of homogeneous condensates (see the previous section). As a result, we have
\begin{eqnarray}
\Omega^{ren}(M,k,k',\Delta)&=&\Omega^{ren}(M,\Delta)\Bigg |_{\nu,\nu_5\to\tilde\nu,\tilde\nu_5}, \label{340}
\end{eqnarray}
where $\Omega^{ren}(M,\Delta)$ is the renormalized TDP (\ref{35}) in the case with homogeneous condensates, i.e. when $k,k'= 0$. However, the TDP (\ref{340}) has several unphysical properties such as (i) the unboundedness from below with respect to the variables $k,k'$. (The unboundedness from below of the TDP (\ref{340}) is evident, e.g., from the expression (\ref{33}) if $\nu_5\to\tilde\nu_5$. In this case the TDP (\ref{33}) behaves as $-\frac{(\nu_5+k')^2}{\pi}$ at $k'\to\infty$.)
(ii) Moreover, one can observe immediately that at $M=0$ and $\Delta=0$ the expression (\ref{340}) for thermodynamic potential does depend on $k$ and  $k^{\prime}$. Bearing in mind the expression (\ref{06}) it is obvious that this is also quite unphysical and we need to change somehow the expression for thermodynamic potential in such a way that this dependence is eliminated.

%%%%%%%%%%%%%%%%%%%%%%%%%%%%%%%%%,
Such unphysical properties of the TDP (\ref{340}) are explained due to the following reason. In the case of spatially homogeneous condensates, all regularization schemes are usually equivalent. However, in the case of spatially inhomogeneous condensate approach  the translational invariance over one or several spatial coordinates is lost. So, the corresponding (spatial) momenta are not conserved. Then, if one uses the momentum-cutoff regularization technique, as in the case of obtaining the expression (\ref{340}), nonphysical (spurious) $k,k'$-dependent terms appear, and the TDP acquires some nonphysical properties (such as the above-mentioned unboundedness from below with respect to $k$, $k'$, etc). In order to obtain a physically relevant TDP (or effective potential), in this case an additional subtraction procedure is usually applied (for details see \cite{miransky,gubina}). For example, if the phase structure of the system is described by only one order parameter (e.g., chiral condensate), then in the spatially inhomogeneous CDW approach the TDP of the system, ${\cal V}(M,k)$, depends on two dynamical quantities, $M$ and $k$ (compare with (\ref{06}), which is for the case under consideration with two inhomogeneous condensates). Then, if renormalized TDP ${\cal V}^{ren}(M,k)$ is obtained in the framework of momentum-cutoff regularization scheme, one should apply to ${\cal V}^{ren}(M,k)$ the following subtraction procedure ${\cal R}_{Mk}$, in order to get a physically relevant TDP ${\cal V}^{phys}(M,k)$ of the system \cite{miransky,gubina}:
\begin{align}%{eqnarray}
&{\cal V}^{phys}(M,k)={\cal R}_{Mk}\Big ({\cal V}^{ren}(M,k)\Big
)\nonumber\\
&\equiv         {\cal V}^{ren}(M,k)-{\cal V}^{ren}(0,k)+{\cal V}^{ren}(0,0).\label{034}
\end{align}%{eqnarray}
(Due to the term ${\cal V}^{ren}(0,k)$ in (\ref{034}) the resulting TDP ${\cal V}^{phys}(M,k)$ becomes bounded from below, whereas the last term ${\cal V}^{ren}(0,0)$ is added there in order the TDP ${\cal V}^{phys}(M,k)$ reproduces at $k=0$ the TDP, obtained in the homogeneous condensate approach.)

On the other hand, if one uses more adequate regularization schemes such as Schwinger proper-time \cite{nakano,heinz} or energy-cutoff regularizations \cite{gkkz,zfk}, etc., such spurious terms do not appear. \footnote{As discussed in the recent papers \cite{gubina,nakano,gkkz,zfk}, an
adequate regularization scheme in the case of spatially inhomogeneous phases
consists in the following: for different quasiparticles the same restriction
on their region of energy values $|P_{01}|,...,|P_{04}|$ should be used in
a regularized thermodynamic potential.}

Since in the paper the momentum-cutoff regularization technique
is used for obtaining the renormalized TDP (\ref{35}), the TDP
(\ref{340}) gets above mentioned nonphysical properties, which should
be eliminated by, e.g., the subtraction operation (\ref{034}) applying
twice, first with respect to the variables $M,k$ and then with respect
to $\Delta,k'$. As a result, we have from Eq. (\ref{340}) the
following physically relevant TDP,
\begin{align}%{eqnarray}
\Omega^{phys}(M,k,k',\Delta)={\cal R}_{\Delta k'}\Big ({\cal
  R}_{Mk}\Big (\Omega^{ren}(M,k,k',\Delta)\Big )\Big ),
\label{42}
\end{align}%{eqnarray}
where $\Omega^{ren}(M,k,k',\Delta)$ is presented in Eq. (\ref{340}). It is
clear from Eq. (\ref{034}) and Eq. (\ref{42}) that
\begin{align}%{eqnarray}
&\Omega^{phys}(M,k,k',\Delta)=\Omega^{ren}(M,k,k',\Delta)-\Omega^{ren}(M,k,k',0)\nonumber\\
&~~~~~+\Omega^{ren}(M,k,0,0)-\Omega^{ren}(0,k,k',\Delta)\nonumber\\
&~~~~~+\Omega^{ren}(0,0,k',\Delta)-\Omega^{ren}(0,k,0,0)\nonumber\\
&-\Omega^{ren}(0,0,k',0)+\Omega^{ren}(0,k,k',0)+\Omega^{ren}(0,0,0,0).
\label{43}
\end{align}%{eqnarray}
It turns out that, as in the case of spatially homogeneous condensates (see Sec. III B), the phase with both spontaneous breaking of chiral and isospin symmetry is absent in the model. So it is enough to study only the projections of the TDP $\Omega^{phys}(M,k,k',\Delta)$ (\ref{43}) on the $M$ and $\Delta$ axes,
\begin{align}\label{44}%{eqnarray}
&\widetilde{F}_{1}(M,k)\equiv\Omega^{phys}(M,k,k',0)=\Omega^{ren}(M,k,0,0)\nonumber\\
&~~~~~-\Omega^{ren}(0,k,0,0)+\Omega^{ren}(0,0,0,0)\nonumber\\
&~~~=F_1(M)\Big |_{\nu\to\tilde\nu}-F_1(0)\Big
|_{\nu\to\tilde\nu}+F_1(0),\\
&\widetilde{F}_{2}(\Delta,k')\equiv\Omega^{phys}(0,k,k',\Delta)\nonumber\\
&=\Omega^{ren}(0,0,k',\Delta)-\Omega^{ren}(0,0,k',0)+\Omega^{ren}(0,0,0,0)\nonumber\\
&=F_2(\Delta)\Big |_{\nu_5\to\tilde\nu_5}-F_2(0)\Big |_{\nu_5\to\tilde\nu_5}+F_2(0),
\label{45}
\end{align}%{eqnarray}
where $F_1(M)$ and $F_2(\Delta)$ are TDPs, which are given in Eq. (\ref{33}) and Eq. (\ref{34}), respectively. Note that the projection $\widetilde{F}_{1}(M,k)$ (the TDP $\widetilde{F}_{2}(\Delta,k')$) does not depend on a wave parameter $k'$ (parameter $k$).

Now, to find the phase structure of the model (1) in the case of inhomogeneous ansatz (\ref{06}) for condensates, it is necessary to determine the global minimum points of the TDPs $\widetilde{F}_1(M,k)$ (\ref{44}) and $\widetilde{F}_2(\Delta,k')$ (\ref{45}) vs $M,k$ and $\Delta,k'$, respectively, and then compare the minimum values of these functions. The result is the GMP of the whole TDP (\ref{42})-(\ref{43}). Investigating the behavior of this GMP vs external parameters $\mu,\nu,\nu_5$, one can establish the phase structure in the approach of spatially inhomogeneous condensates (\ref{06}).

There could be several phases in the model (1). The first one is the symmetric phase, which corresponds to the global minimum point $(M_0,k_{0},k^{\prime}_{0},\Delta_0)$ of the TDP (\ref{43}) with zero gaps $M_0=0,\Delta_0=0$ and zero values of the wavevectors $k_{0}=0,k^{\prime}_{0}=0$. In the chiral symmetry breaking phase, homogeneous or inhomogeneous, the TDP (\ref{43}) reaches the least value at the global minimum point with $M_0\ne 0,\Delta_0=0, k^{\prime}_{0}=0$ and  $k_{0}=0$ or $k_{0}\neq 0$, respectively. Finally, in the charged pion condensation phase, homogeneous or inhomogeneous, the global minimum of the TDP lies at the point with $M_0=0,\Delta_0\ne 0, k_{0}=0$ and  $k^{\prime}_{0}=0$ or $k^{\prime}_{0}\neq 0$, respectively. (Notice that in the most general case the coordinates of the global minimum point, i.e. the gaps $M_0$, $\Delta_0$ and the wavevectors $k_{0}$, $k^{\prime}_{0}$, depend on chemical potentials.)
Since in our consideration the quark number density $n_q$ is the most important physical feature of the ground state, we present here the ways how expressions for $n_q$ can be found in different phases. Recall that in the general case this quantity is defined by the relation
\begin{eqnarray}
n_q=-\frac{\partial\Omega^{phys}(M_0,k_{0},k^{\prime}_0,\Delta_0)}{\partial\mu}. \label{37}
\end{eqnarray}
Hence, in the chiral symmetry breaking phase we have from Eq. (\ref{37}) and Eq. (\ref{43}) that
\begin{align}%{eqnarray}
n_q\bigg |_{CSB}&=&-\frac{\partial\Omega^{phys}(M_0,k_{0},k^{\prime}_0,\Delta_0=0)}{\partial\mu}=-\frac{\partial
\widetilde{F}_1(M_0,k_{0})}{\partial\mu},
\label{38}
\end{align}%{eqnarray}
where the quantity $\widetilde F_1(M,k)$ is given in Eq. (\ref{44}), whereas the particle density in the charged pion condensation phase looks like
\begin{align}%{eqnarray}
n_q\bigg |_{PC}=-\frac{\partial\tilde{\Omega}^{phys}(M_0=0,k_{0},k^{\prime}_{0},\Delta_0)}{\partial\mu}=-\frac{\partial
\widetilde{F}_2(\Delta_0,k^{\prime}_{0})}{\partial\mu},
\end{align}%{eqnarray}
where the quantity $\widetilde F_2(\Delta,k')$ is given in Eq. (\ref{45}).

 \subsection{ Phase diagrams and duality property of the model }
\begin{figure*}
%----figure 3
\includegraphics[width=0.45\textwidth]{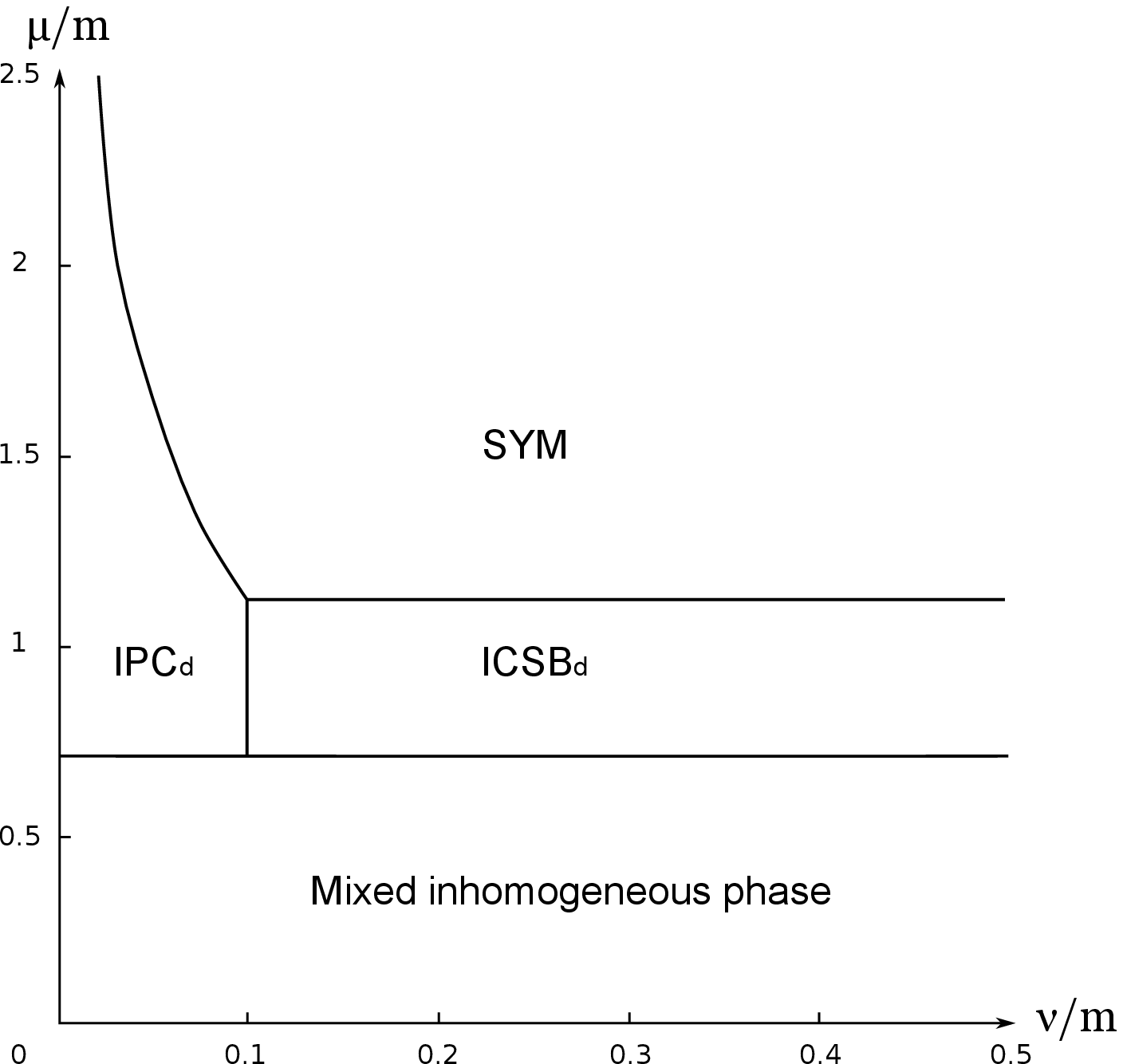}
\hfill
\includegraphics[width=0.45\textwidth]{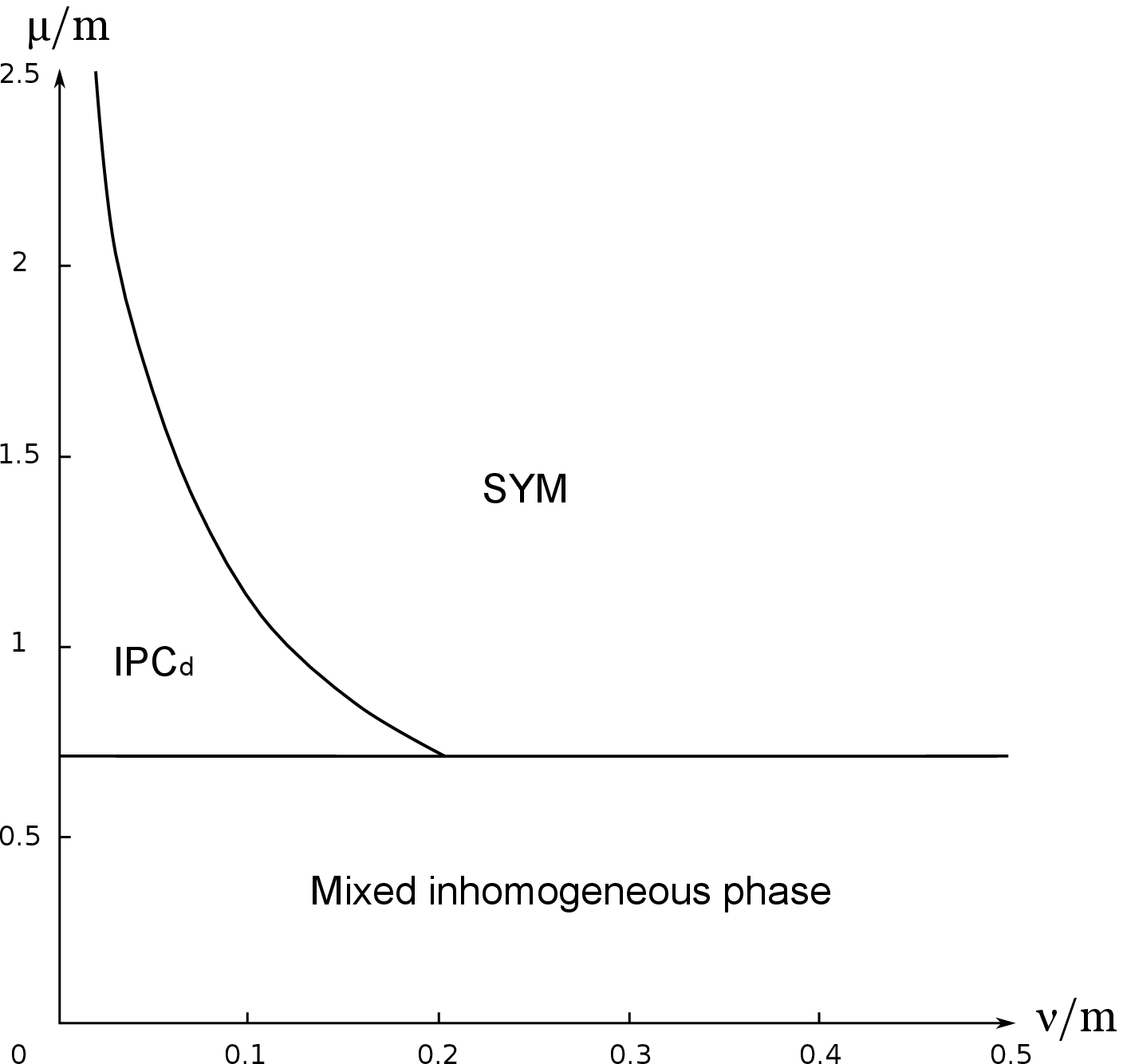}\\
\caption{{\bf The case of spatially inhomogeneous condensates}: The $(\nu,\mu)$-phase portrait of the model at  $\nu_5=0.1m$ (left figure) and at  $\nu_5\ge 0.2m$ (right figure). Here SYM denotes the symmetrical phase, in the IPCd region a global minimum of the TDP corresponds to the inhomogeneous charged pion condensation phase with nonzero quark number density. Other notations are the same as in Fig. 2. }
\end{figure*}
As it was mentioned in the previous section, in order to obtain the most general $(\mu,\nu,\nu_{5})$-phase portrait (diagram) of the model in the framework of the spatially inhomogeneous ansatz (\ref{06}) for condensates, we have to study (numerically) the behavior of the TDP (\ref{43}) global minimum point $(M_0,k_{0},k^{\prime}_{0},\Delta_0)$ vs chemical potentials. However, to simplify the task, it is very convenient to consider different cross-sections of this diagram by the planes of the form $\nu= const$, $\nu_5= const$ and $\mu= const$. These particular phase portraits will help us to form an understanding of the structure of the most general $(\mu,\nu,\nu_{5})$-phase portrait of the model. Moreover, by this way one can use very efficiently the duality symmetry (\ref{16}) of the model TDP (\ref{9}), which allows to predict a phase structure in different regions of the $(\mu,\nu,\nu_{5})$ plane.

\subsubsection{Different $(\mu,\nu)$-phase diagrams}
\label{mixed}

First, we consider the $(\mu,\nu)$-phase portraits of the model at several typical values of $\nu_5$. Recall that in Ref. \cite{gubina} this phase portrait was investigated at $\nu_5=0$. In addition, it was supposed in \cite{gubina} that in the ground state of the system only chiral condensate is spatially inhomogeneous (as CDW), but pion condensate is a homogeneous one, i.e. from the very beginning  the ansatz (\ref{06}) was used there with $k\ne 0$, $k'=0$ for condensates. In this case the $(\mu,\nu)$-phase diagram has a rather trivial form. Namely, for arbitrary $\nu >0$ there is a homogeneous charged pion condensation (PC) phase if $\mu < m/\sqrt{2}$, and at larger values of $\mu$ one can observe the inhomogeneous chiral symmetry breaking (ICSB) phase ($m$ is a massive parameter presented in Eq. (\ref{30})).

Our analysis shows that at infinitesimal values of $\nu_5$, i.e. at $\nu_5=0_+$, this phase portrait is changed if both the chiral and charged pion condensates are spatially inhomogeneous in the form of Eq. (\ref{06}) (see Fig. 2). Indeed, while at $\mu> m/\sqrt{2}$ one can see there the same ICSBd phase (here and below the additional symbol ''d`` means that quark number density in the phase is nonzero), at lower values of $\mu$ there is a region which is called ''Mixed inhomogeneous phase``. It turns out that for each point $(\mu,\nu)$ belonging to this region the TDP (\ref{30}) has two degenerate global minima, first of them, i.e. the point  of the form $(M_0=m,k_{0}=-\nu,k^{\prime}_{0}=0,\Delta_0=0)$, corresponds to ICSB phase, the second -- the point of the form $(M_0=0,k_{0}=0,k^{\prime}_{0}=-\nu_5,\Delta_0=m)$ -- to inhomogeneous charged pion condensation (IPC) phase. The degeneracy of these ground states means that for arbitrary fixed values of chemical potentials $\mu$ and $\nu$ from this region in the space, filled with ICSB (or chiral density wave) phase, a bubble of the IPC phase (and vice versa) can be created, i.e. one can observe in  space the mixture (or coexistence) of these two phases. Note also that in mixed inhomogeneous phase of Fig. 2, etc the quark number density is zero.
\begin{figure*}
%----figure 4,5
\includegraphics[width=0.45\textwidth]{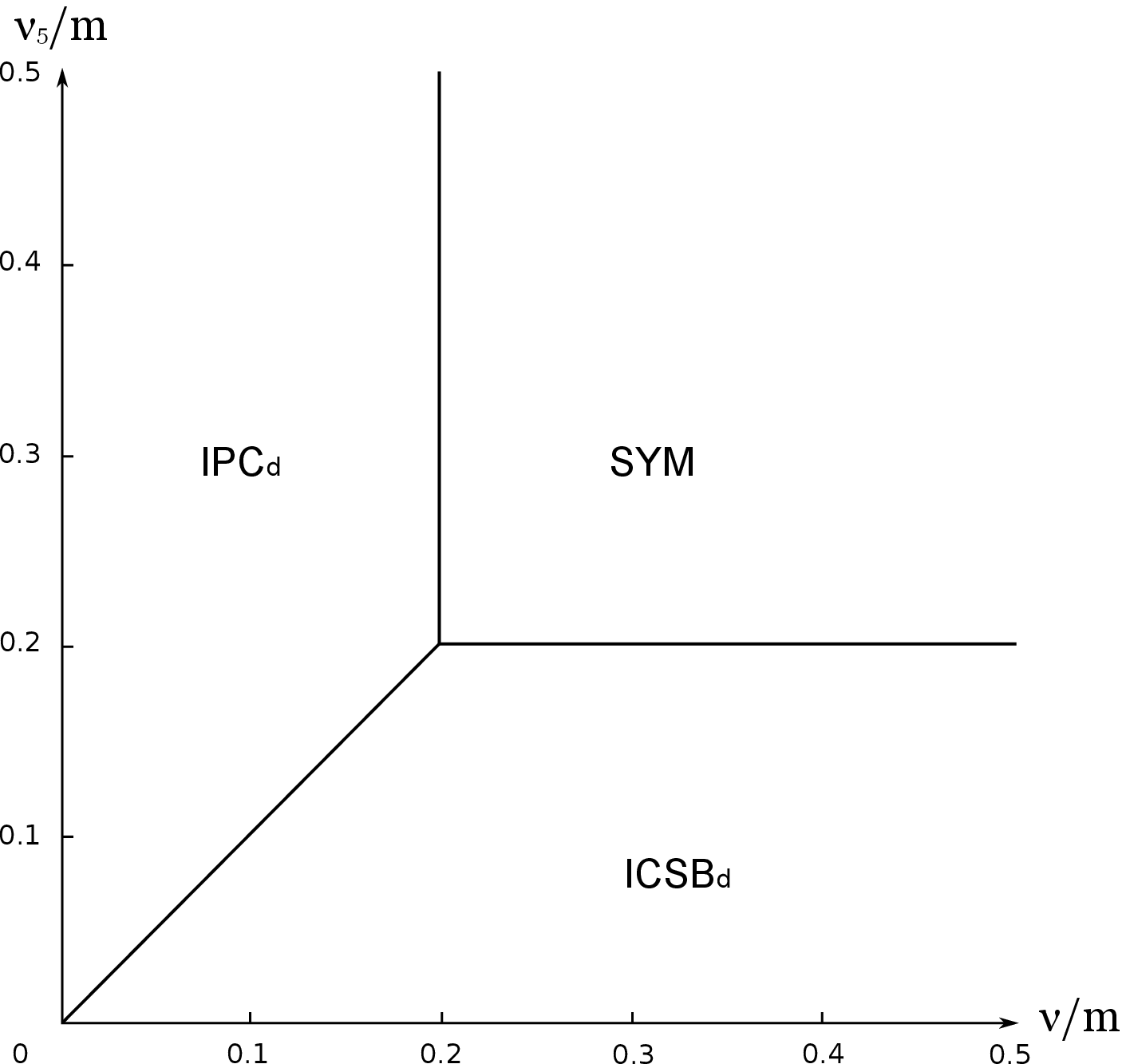}
\hfill
\includegraphics[width=0.45\textwidth]{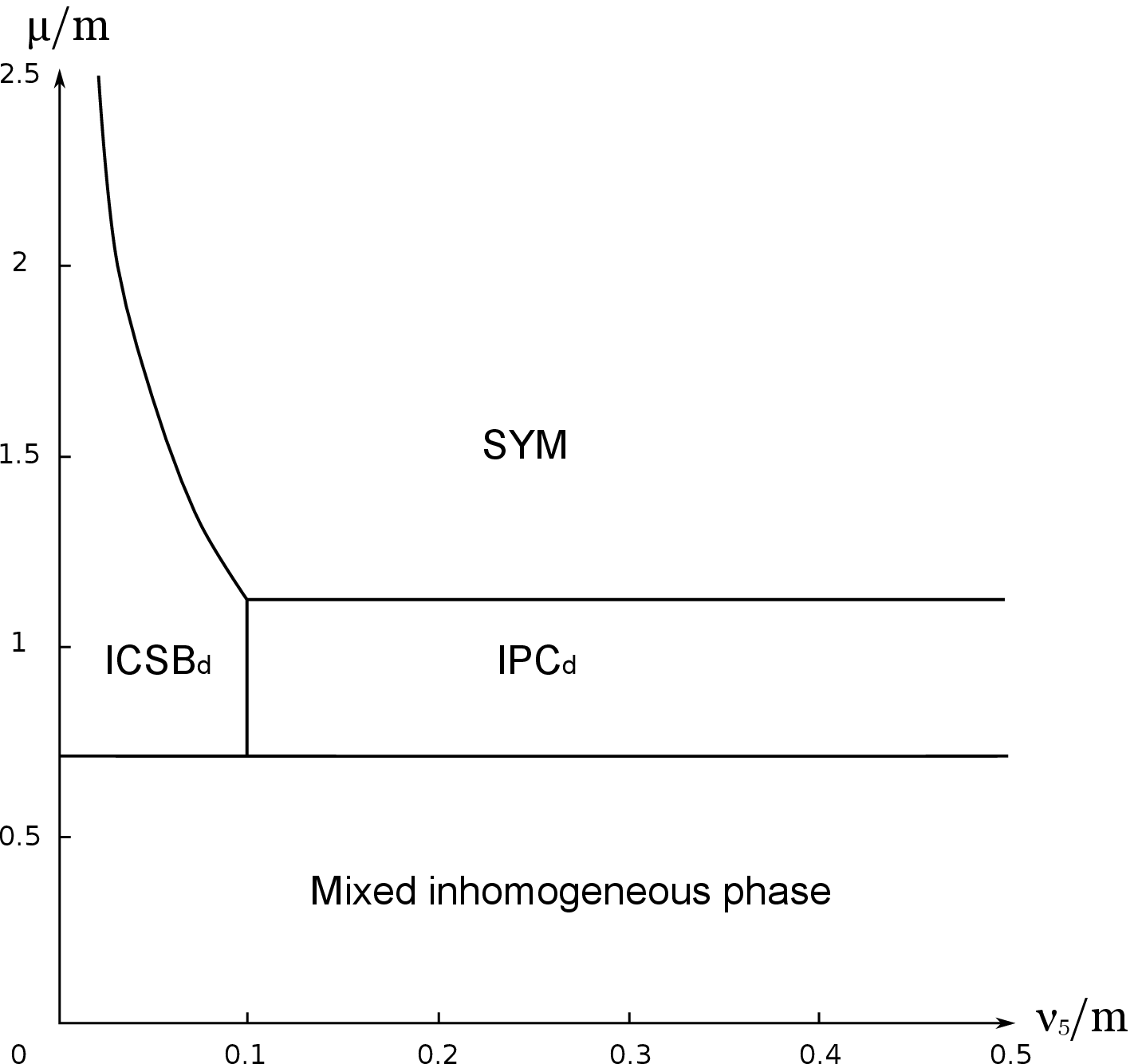}\\
\parbox[t]{0.45\textwidth}{\caption{{\bf The case of spatially inhomogeneous condensates}: The $(\nu,\nu_5)$-phase portrait of the model at  $\mu=0.75m$.
All notations are described in Figs 2, 3.}}\hfill
\parbox[t]{0.45\textwidth}{\caption{{\bf The case of spatially inhomogeneous condensates}: The $(\nu_5,\mu)$-phase portrait of the model at  $\nu=0.1m$. All notations are described in Figs 2, 3.}}
\end{figure*}

The structure of $(\mu,\nu)$-phase diagrams at other fixed values of the chiral chemical potential $\nu_5$ can be easily understood from the phase portraits of Figs 3, where $(\mu,\nu)$-phase diagrams are presented for two qualitatively different values of $\nu_5$. It clear from the figure that at each finite $\nu_5>0$ the $(\mu,\nu)$-phase diagram contains the inhomogeneous charged pion condensation phase with nonzero quark number density (IPCd). Moreover, the greater $\nu_5$, the smaller the size of the ICSBd phase, which disappears from a $(\mu,\nu)$-phase portrait at $\nu_5\ge 0.2m$. Hence, in
%%%%%%%%%%%%%%%%%%%%%%%%%%%%%%%%%%%%%%%%%%5
 the framework of the initial NJL$_2$ model, the chiral chemical potential $\nu_5$ serves as a factor, which promotes the charged pion condensation
%%%%%%%%%%%%%%%%%%%%%%%%%%%
 phenomenon in dense quark matter (it is the IPCd phase in all figures).%%%
%%%%%%%%%%%%%%%%%%%%%%%%%%%%%%5
%%%%%%%%%%%%%%%%%%%%%%%%%%%%%%%%%%the
%
%Then it depends on the fact which phase is realized.
%Let us take for example the transition form mixed inhomogeneous phase to  IPC.
%If the system is in the IV ICSB and IPC degenerate phase then it choses one of the equivalent phases.
%If in mixed inhomogeneous phase IPC is realized then there is no phase transition at all.
%If in mixed inhomogeneous phase ICSB then  it is first order phase transition.
%%%%%%%%%%%%%%%%%%%%%%%%%%%%%%%%%%%%%%%

\subsubsection{Other phase diagrams and the role of duality }

Before presenting the $(\mu,\nu_5)$- and $(\nu,\nu_5)$-phase diagrams at fixed values of $\nu$ and $\mu$, respectively, and before obtaining of the most general $(\nu,\nu_5,\mu)$-phase portrait of the model, let us discuss the role and influence of the duality invariance  (\ref{16}) of the model TDP (\ref{9})-(\ref{43}) on the phase structure.

Suppose that at some fixed particular values of chemical potentials $\mu$, $\nu=A$ and $\nu_5=B$ the global minimum of the TDP (\ref{43}) lies at the
point, e.g., $(M=M_0\ne 0,k=k_0,k'=0,\Delta=0)$. It means that for such fixed values of the chemical potentials the chiral symmetry breaking (CSB) phase is
realized in the model (it is  homogeneous if $k_0=0$ or inhomogeneous if $k_0\ne 0$). Then it follows from the duality invariance of the TDP (\ref{9}) (or (\ref{43})) with respect to the transformation ${\cal D}$ (\ref{16}) that at permuted chemical potential values (i.e. at $\nu=B$ and $\nu_5=A$ and intact value of $\mu$) the global minimum of the TDP $\Omega^{phys}(M,k,k',\Delta)$ is arranged at the point $(M=0,k=0,k'=k_0,\Delta=M_0)$, which corresponds to the charged, homogeneous or inhomogeneous, PC phase (and vice versa). This is the so-called duality correspondence between CSB and charged PC phases in the framework of the model under consideration.

Hence, the knowledge of a phase of the model (1) at some fixed values of external free model parameters $\mu,\nu,\nu_5$ is sufficient  to understand what phase (we call it a dually conjugated) is realized at rearranged values of external parameters, $\nu\leftrightarrow\nu_5$, at fixed $\mu$. Moreover, different physical parameters such as condensates, densities, etc, which characterize both the
initial phase and the dually conjugated one, are connected by the duality transformation ${\cal D}$. For example, the chiral condensate of the initial CSB phase at some fixed $\mu,\nu,\nu_5$ is equal to the charged-pion condensate of the dually conjugated charged PC phase, in which one should perform the replacement  $\nu\leftrightarrow\nu_5$. Knowing the particle density $n_q(\nu,\nu_{5})$ of the initial CSB phase as a function of chemical potentials $\nu,\nu_{5}$, one can find the particle density in the dually conjugated charged PC phase by interchanging $\nu$ and $\nu_{5}$ in the expression $n_q(\nu,\nu_{5})$, etc.

The duality transformation ${\cal D}$ of the  TDP can also be applied to an arbitrary phase portrait of the model. In particular, it is clear that if we have a most general $(\nu,\nu_5,\mu)$-phase portrait, i.e. the one-to-one correspondence between any point $(\nu,\nu_5,\mu)$ of the three-dimensional space of chemical potentials and possible model phases (CSB, charged PC and symmetric phase), then under the duality transformation (which is understood as a renaming both of the diagram axes, i.e. $\nu\leftrightarrow\nu_5$, and phases, i.e.  CSB$\leftrightarrow$charged PC) this phase portrait is mapped to itself, i.e. the most general $(\nu,\nu_5,\mu)$-phase portrait is self-dual. Furthermore, the self-duality of the $(\nu,\nu_5,\mu)$-phase portrait means that in the three-dimensional $(\nu,\nu_5,\mu)$ space the regions of the CSB and charged PC phases (both homogeneous and inhomogeneous) are arranged mirror-symmetrically with respect to the plane $\nu=\nu_5$ of this space.
\begin{figure*}
%----figure 6,7
\includegraphics[width=0.45\textwidth]{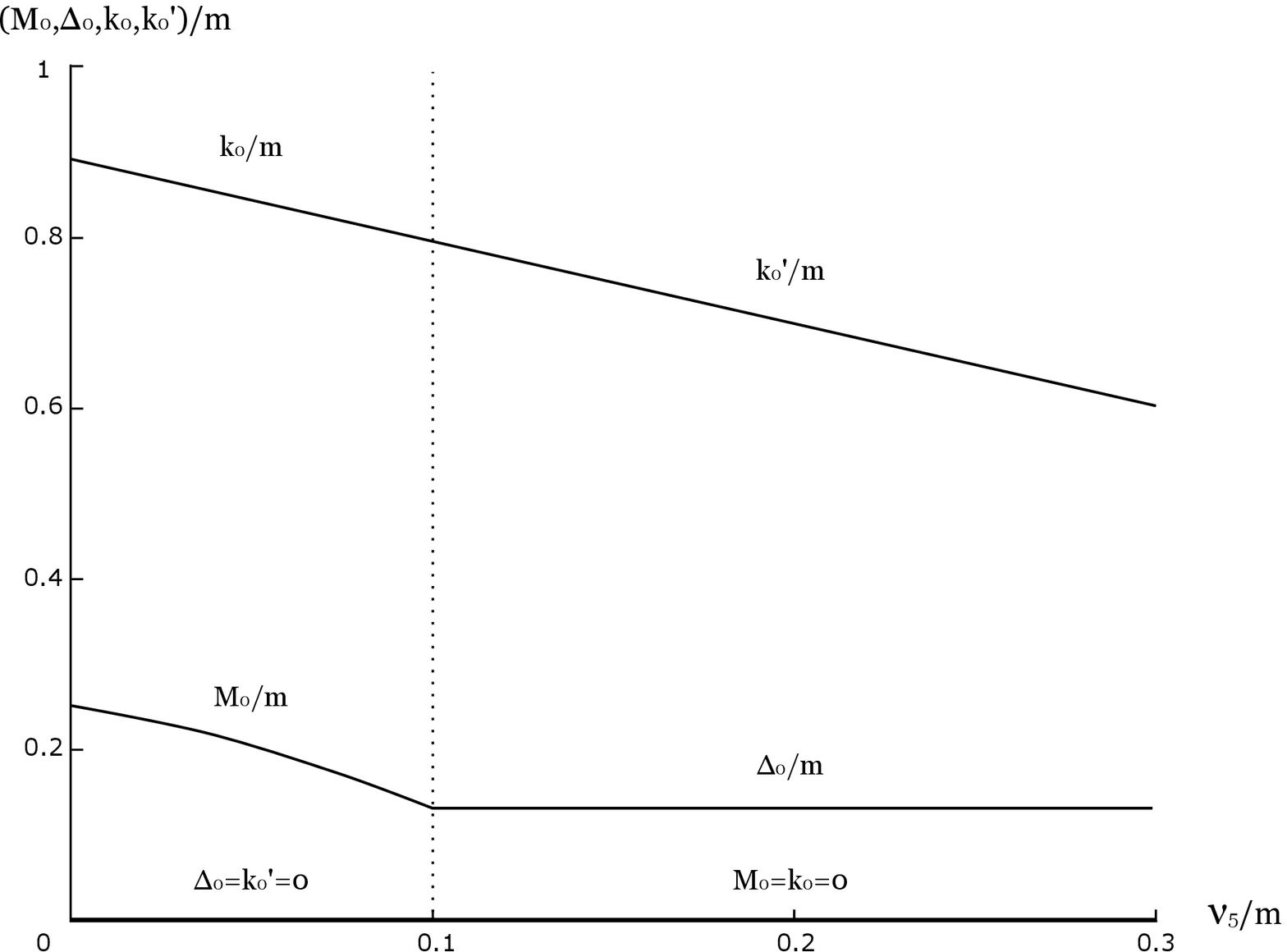}
\hfill
\includegraphics[width=0.45\textwidth]{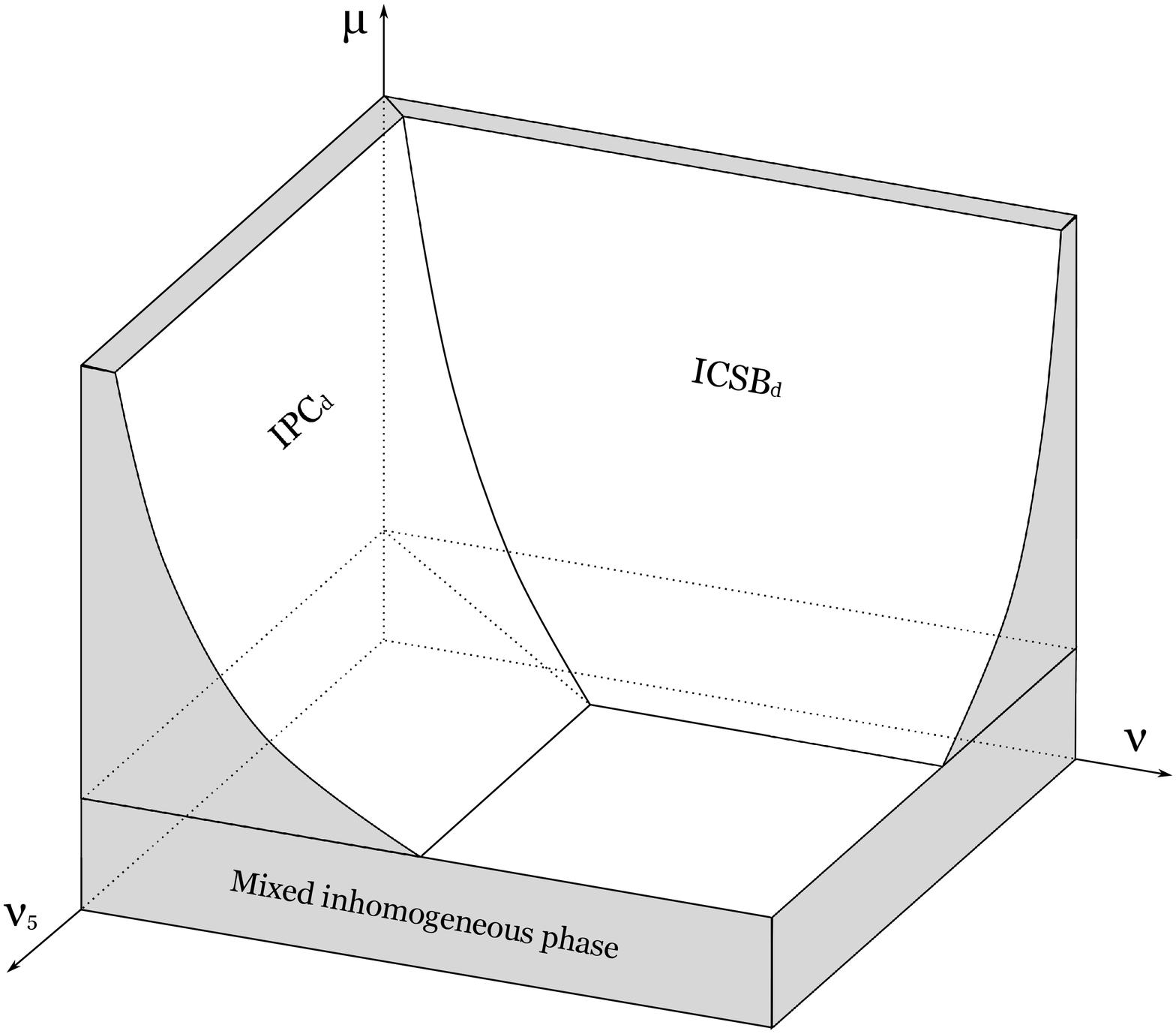}\\
\parbox[t]{0.45\textwidth}{\caption{{\bf The case of spatially inhomogeneous condensates}: The behavior of the coordinates $M_0,k_0,k'_0,\Delta_0$ of the GMP of the TDP (\ref{43}) as functions of $\nu_5$ for fixed $\mu=m$ and $\nu=0.1m$.}}\hfill
\parbox[t]{0.45\textwidth}{\caption{Schematic representation of the $(\nu_5,\nu,\mu)$-phase portrait of the model in the case of spatially inhomogeneous condensates. The notations are the same as in Figs 2, 3. The points which are outside IPCd-, ICSBd- and ''Mixed inhomogeneous phase''-regions of the diagram correspond to the symmetric phase of the model.}}
\end{figure*}

It follows from these  rather general inferences that at arbitrary fixed $\mu$ the $(\nu,\nu_5)$-phase diagram of the model is also self-dual, i.e. its CSB and charged PC (homogeneous or inhomogeneous) phases should lie mirror-symmetrically with respect to the line $\nu=\nu_5$. These conclusions are supported by Fig. 4, where we present a typical self-dual $(\nu,\nu_5)$-phase portrait of the model for the case $\mu> m/\sqrt{2}\approx 0.71m$.
%It is easily seen from Fig. 4, where we present a typical self-dual $(\nu,\nu_5)$-phase portrait for the case $\mu> m/\sqrt{2}\approx 0.71m$, which supports our conclusions. 
If $\mu< m/\sqrt{2}$, then the $(\nu,\nu_5)$-phase portrait is even simpler because at each point of it the ``Mixed inhomogeneous phase`` is realized. The features of the structure of this phase were already discussed in details in the previous section \ref{mixed}. So it is clear that both the ''Mixed inhomogeneous phase`` itself and the $(\nu,\nu_5)$-phase portrait of the model at fixed $\mu< m/\sqrt{2}$  are also  self-dual.

Now let us show how to construct the $(\nu_5,\mu)$-phase diagram of the model at arbitrary fixed value $\nu=A$. Of course, in this case one can fulfill a numerical investigation of the TDP (\ref{43}). However, a simpler way (which is due to the dual invariance (\ref{16}) of the TDP) is to perform the dual transformation of the $(\nu,\mu)$-phase diagram at the corresponding fixed value $\nu_5=A$. For example, to find the $(\nu_5,\mu)$-phase diagram at $\nu=0.1m$ we should start from the $(\nu,\mu)$-diagram at fixed $\nu_5=0.1m$ of Fig. 3 (left panel) and make the simplest replacement in the notations of this figure: $\nu\leftrightarrow\nu_5$, IPCd$\leftrightarrow$ICSBd. (Note, the symmetrical and mixed inhomogeneous phases are intact under the dual transformation.) As a result of this mapping, we obtain the phase diagram of Fig. 5. In a similar way one can dually transform Fig. 2 and Fig. 3 (right panel)  in order to find the $(\nu_5,\mu)$-phase diagrams at $\nu=0_+$ and $\nu\ge 0.2m$, respectively, etc. 

%%%%%%%%%
The behaviour of different order parameters, such as chiral $M_0$ and charged PC $\Delta_0$ condensates as well as corresponding wave vectors $k_0$ and $k'_0$ (note that the point $(M_0,k_0,k'_0,\Delta_0)$ is the GMP of the TDP (\ref{43})), as functions of the chemical potential $\nu_5$ at fixed values of $\mu=m$ and $\nu=0.1m$ is presented in Fig. 6. It is clear from the figure that in the critical point $\nu_5=0.1m$ there is a phase transition in the system from ICSBd phase (where the GMP has the form $(M_0\ne 0,k_0\ne 0,k'_0=0,\Delta_0=0)$) to IPCd phase (where the GMP looks like $(M_0= 0,k_0= 0,k'_0\ne 0,\Delta_0\ne 0)$). Since in this point a GMP of the system changes its location by a jump, we conclude that it is a phase transition of the first order.

Now, taking into account the particular phase diagrams of Figs 2--5, it is possible to represent schematically the most general phase portrait of the model in the space of chemical potentials $\nu,\nu_5,\mu$ (see Fig. 7). As is easily seen from this figure, the charged pion condensation and chiral symmetry breaking phases are arranged mirror-symmetrically with respect to the plane $\nu=\nu_5$, i.e. the phase diagram is self-dual. Moreover, it supports the above conclusion: the charged PC phenomenon can be realized in chirally asymmetric quark matter with nonzero baryon density.

Finally, few words about the order of phase transitions between phases depicted in Fig. 7. Numerical analysis of the TDP (\ref{43}) gives us the form of its GMP $(M_0,k_0,k'_0,\Delta_0)$ and especially the behavior of the GMP vs chemical potentials on the boundaries between phases. Since on the boundary between ICSBd and IPCd phase the GMP changes its position in the $(M,k,k',\Delta)$ space by a jump, we conclude that on this boundary there are first order phase transitions. (The situation is well illustrated by Fig. 6, where for particular values of chemical potentials just a first order phase transition occurs between ICSBd and IPCd phases.) However, numerical investigations show that on the boundary between ICSBd and symmetrical as well as between IPCd and symmetrical phases there occurs a second order phase transition. In our opinion, when one talks about the transitions from the region ''Mixed  inhomogeneous phase'' to other phases, it does  not make sense to talk about the order of the phase transition or even about the phase transition itself. The reason is the fact that in ``Mixed  inhomogeneous phase`` the system has two equivalent minima corresponding to IPC or ICSB phases, and we do not know which one is realized at the moment. 
%\begin{figure}
%----figure 6
%\includegraphics[width=0.45\textwidth]{fig6.eps}
%\hfill
%\includegraphics[width=0.45\textwidth]{fig3b.eps}\\
%\caption{Schematic representation of the $(\nu_5,\nu,\mu)$-phase portrait of the model in the case of spatially inhomogeneous condensates. The notations are the same as in Figs 2, 3. The points which are outside IPCd-, ICSBd- and ''Mixed inhomogeneous phase''-regions of the diagram correspond to the symmetric phase of the model.}
%\end{figure}

\section{Summary and conclusions}

In this paper the phase structure of the NJL$_2$ model (1) with two
quark flavors is investigated in the large-$N_c$ limit in the presence
of baryon $\mu_B$, isospin $\mu_I$ and chiral isospin $\mu_{I5}$ chemical
potentials. Moreover, we take into account the possibility of existence of spatially inhomogeneous condensates which are assumed to have the form of
a chiral density wave for chiral condensate and a single plane wave for charged pion one (see in Eq. (\ref{06})).

Recall that for the particular case with $\mu_{I5}=0$ and with account
only for chiral density wave, the problem was solved earlier in Refs
\cite{gubina,andersen2}, where it was shown that the toy model (1)
does not predict a charged PC phase of dense and isotopically asymmetric quark matter. If $\mu_{I5}\neq 0$ and both condensates are spatially homogeneous, this model was considered in Ref. \cite{ekk}, where it was shown  that $\mu_{I5}$ promotes charged PC phase with nonzero baryon density. The most general $(\nu,\nu_5,\mu)$-phase diagram of the model in this case is depicted schematically in Fig. 1 of the present paper, where this phase is
denoted as PCd. In contrast to Ref. \cite{ekk}, our present consideration of the model (1) is mainly devoted to study of the properties of chirally ($\mu_{I5}\ne 0$) and isotopically ($\mu_{I}\ne 0$) asymmetric  dense ($\mu_{B}\ne 0$) quark matter with inhomogeneous condensates of the form (\ref{06}).

Let us summarize some of the most interesting results obtained.

1) We proved that the main conclusion made in Ref. \cite{ekk} for the model (1)  under the assumption that all condensates are homogeneous, i.e., that charged PC phase with nonzero baryon density can be realized in chirally asymmetric dense quark matter, holds also for the case of spatially inhomogeneous condensates. Furthermore, in the last case this phase is realized for larger region of a phase diagram and hence for broader range of $\mu_{I5}$ (compare Fig. 1 and Fig. 7). In particular, in the inhomogeneous case the dense PC phase is realized at any nonzero $\mu_{I5}$ in contrast to homogeneous case, where it is realized only for rather large values of  $\mu_{I5}$ (larger than some particular value). This means that in dense quark matter charged pion condensation takes place even at small chiral asymmetry.

2) In this model inhomogeneous condensates are quite favoured compared to
homogeneous condensates. For nonzero values of the chiral isospin chemical
potential $\mu_{I5}$ and isospin chemical potential $\mu_{I}$, all phases at the phase diagram are inhomogeneous or symmetric ones.

3) We demonstrated in the framework of the NJL$_2$ model (1), that in the inhomogeneous case duality  correspondence between CSB and charged PC phenomenon takes place in the leading order of the large-$N_c$ approximation.

Finally, the main result of this paper is that the chemical potential $\mu_{I5}$ generates charged pion condensation in dense quark matter. This effect is realized both in spatially homogeneous and inhomogeneous approaches for condensates, but in the inhomogeneous case it is even enhanced.  

% In PC phase  parity is broken. In the inhomogeneous case (in IPC phase) charge conjugation $C$ symmetry is broken as well. 
%There is a number of works that explored the idea that QCD breaks %parity
% P  and CP symmetry  at high temperatures and densities. \cite{espr1,kharzpisart,AndAndE,KharZ,AndAndEP,ChFLoST,Es,Vol}
%First, analytical studies with effective models suggest that QCD may break parity. Second, P and CP odd bubbles may appear in a
%finite volume due to local large topological fluctuations in a hot medium dense systems. And it is expected that these P-odd bubbles may be produced in heavy ion collisions.
%%%%%%%%%%%%%%%%%%%%%%%%%
After our work was finished, we found a recent paper \cite{kaw}, where 
it was discussed a possibility for charged pions to act as a probe for measuring the strong CP violation in chirally imbalance matter. Since in the charged PC phase the isotopic density is nonzero, which also implies the possibility of a nonzero density of charged pions, we guess that our results support this mechanism for detecting the CP violation in chirally asymmetric baryon matter. Moreover, we believe that our analysis may shed some new light on other physical effects in chirally and isotopically asymmetric dense quark matter in the case of realistic (3+1)-dimensional spacetime. 
It is applied primarily to heavy-ion colliding systems, where the external magnetic field can reach high values, and spacetime is effectively (1+1)-dimensional.

\appendix

\section{Evaluation of the roots of the polynomial $P_4(p_0)$ (\ref{91}) at $k=k'=0$ }
\label{ApB}
\subsection{General consideration}
\vspace{-0.4cm}

At $k=k'=0$ it is very convenient to present the fourth-order polynomial (\ref{91}) of the variable $\eta\equiv p_0+\mu$  as a product of two second-order polynomials (this way is proposed in \cite{Birkhoff}), i.e. we assume that
\begin{align}%{eqnarray}
&\eta^4-2a \eta^2-b \eta +c = (\eta^2 + r\eta + q)(\eta^2 - r\eta + s)\nonumber\\
&=\left [\left (\eta+\frac r2\right )^2+q-\frac{r^2}{4}\right ]\left [\left
(\eta-\frac r2\right )^2+s-\frac{r^2}{4}\right ]\nonumber\\
&\equiv(\eta-\eta_{1})(\eta-\eta_{2})(\eta-\eta_{3})(\eta-\eta_{4}),\label{B3}
\end{align}%{eqnarray}
where $r$, $q$ and $s$ are some real valued quantities, such that (see the relations (\ref{91})):
\begin{align}%{eqnarray}
&-2a~\equiv-2(M^2+\Delta^2+p_1^2+\nu^2+\nu_{5}^2)= s+q-r^2;~~
\nonumber\\
&-b\equiv -8p_1\nu\nu_{5}= rs-qr;~~c~\equiv a^2-4p_1^2(\nu^2+\nu_5^2)\nonumber\\
&~~~~~-4M^2\nu^2-4\Delta^2\nu_5^2-4\nu^2\nu_5^2=sq.
\label{B4}
\end{align}%{eqnarray}
In the most general case, i.e. at $M\ge 0$, $\Delta\ge 0$, $\nu\ge 0$,  
$\nu_5\ge 0$ and arbitrary values of $p_1$, one can solve the system of equations (\ref{B4}) with respect to $q,s,r$ and find
\begin{align}%{eqnarray}
&q=\frac 12 \left (-2a +R+\frac{b}{\sqrt{R}}\right ),~~ s=\frac 12 \left
(-2a +R-\frac{b}{\sqrt{R}}\right ),~~\nonumber\\
&r=\sqrt{R},\label{B5}%~~~~~B5
\end{align}%{eqnarray}
where $R$ is an arbitrary positive real solution of the equation
\begin{eqnarray}
\label{B6} X^3 + AX=BX^2 + C%~~~~~~~~B6
\end{eqnarray}
with respect to a variable $X$, and
\begin{align}%{eqnarray}
&A=4a^2-4c=16\Big[\nu_5^2\Delta^2+M^2\nu^2+
\nu_5^2\nu^2+p_1^2(\nu^2+\nu_5^2)\Big],\nonumber\\
&B=4a =4(M^2+\Delta^2+\nu^2+\nu_5^2+p_1^2),\nonumber\\
&C=b^2=(8\nu_5\nu p_1)^2. \label{B7}
\end{align}%{eqnarray}
Finding (numerically) the quantities $q$, $s$ and $r$, it is possible to obtain from Eq. (\ref{B3}) the roots $\eta_i$:
\begin{align}%{eqnarray}
&\eta_{1}=-\frac{r}{2}+\sqrt{\frac{r^2}{4}-q},~~\eta_{2}=\frac{r}{2}+
\sqrt{\frac{r^2}{4}-s},~~\nonumber\\
&\eta_{3}=-\frac{r}{2}-\sqrt{\frac{r^2}{4}-q},~~\eta_{4}=\frac{r}{2}-\sqrt{\frac{r^2}{4}-s}.\label{B41}%~~~~B41
\end{align}%{eqnarray}
Numerical investigation shows that in the most general case the discriminant of the third-order algebraic
equation (\ref{B6}), i.e. the quantity $18ABC-4B^3C+A^2B^2-4A^3-27C^2$, is
always nonnegative. So the equation (\ref{B6}) vs $X$ has three real
solutions $R_1,R_2$ and $R_3$ (this fact is presented in \cite{Birkhoff}).
Moreover, since the coefficients $A$, $B$ and $C$ (\ref{B7}) are nonnegative, it is clear, due to a form of the equation (\ref{B6}), that all its roots $R_1$, $R_2$ and $R_3$ are also nonnegative quantities (usually, they are positive and different). So we are free to choose the quantity $R$ from (\ref{B5}) as one of the positive solutions $R_1$, $R_2$ or $R_3$. In each case, i.e. for $R=R_1$, $R=R_2$, or $R=R_3$, we will obtain the same set of the roots (\ref{B41}) (possibly rearranged), which depends only on $\nu$, $\nu_5$, $M$, $\Delta$ and $p_1$, and does not depend on the choice of $R$. Due to the relations (\ref{B3})-(\ref{B41}), one can find numerically (at fixed values of  $\mu$, $\nu$, $\nu_5$, $M$, $\Delta$ and $p_1$) the roots $\eta_i=p_{0i}+\mu$ (\ref{B41}) and, as a result, investigate numerically the TDP (\ref{28}). It is clear also from Eqs
(\ref{B3})-(\ref{B41}) that the roots $\eta_i$ are even functions vs $p_1$. So in all improper $p_1$ integrals, which include quasiparticle energies $p_{0i}$ (see, e.g., the integral in Eq. (\ref{28})), we can confine ourselves by integration over nonnegative values of $p_1$ (up to a factor 2).

On the basis of the relations (\ref{B3})-(\ref{B41}) let us consider the asymptotic behavior of the quasiparticle energies $p_{0i}$ at $p_1\to\infty$. First of all, we start from the asymptotic analysis of the roots $R_{1,2,3}$ of the equation (\ref{B6}) at $p_1\to\infty$,
\begin{align}%{eqnarray}
\label{B701}
&R_1=4\nu^2-\frac{4\Delta^2\nu^2}{p_1^2}
+{\cal O}\big (1/p_1^4\big ),\\
\label{B7001}
&R_2=4\nu_5^2-\frac{4M^2\nu_5^2}{p_1^2}
+{\cal O}\big (1/p_1^4\big ),\\
&R_3=4p_1^2+4(M^2+\Delta^2)+\frac{4(\nu_5^2M^2+\nu^2\Delta^2)}{p_1^2}
+{\cal O}\big (1/p_1^4\big ). \label{B71}
\end{align}%{eqnarray}
It is clear from these relations that $R_3$ is invariant, whereas
$R_1\leftrightarrow R_2$ under the duality transformation (\ref{16}). Then, using for example $R_3$ (\ref{B71}) as the quantity $R$ in Eqs. (\ref{B5}) and (\ref{B41}), one can get the asymptotics of the quasiparticle energies $p_{0i}\equiv \eta_i-\mu$ at $p_1\to\infty$,
\begin{align}%{eqnarray}
&p_{01}=-|p_1|-\mu+|\nu_5-\nu|-\frac{\Delta^2+M^2}{2|p_1|} +{\cal O}\big
(1/p_1^2\big ),~~ \nonumber\\
&p_{02}=|p_1|-\mu+\nu_5+\nu+\frac{\Delta^2+M^2}{2|p_1|}
+{\cal O}\big (1/p_1^2\big ),\nonumber\\
&p_{03}=-|p_1|-\mu-|\nu_5-\nu|-\frac{\Delta^2+M^2}{2|p_1|} +{\cal O}\big
(1/p_1^2\big ),~~\nonumber\\
& p_{04}=|p_1|-\mu-\nu_5-\nu+\frac{\Delta^2+M^2}{2|p_1|}
+{\cal O}\big (1/p_1^2\big ).\label{B26}
\end{align}%{eqnarray}
Finally, it follows from (\ref{B26}) that at $p_1\to\infty$
\begin{align}%{eqnarray}
&|p_{01}|+|p_{02}|+|p_{03}|+|p_{04}|=4|p_1|+\frac{2(\Delta^2+M^2)}{|p_1|}\nonumber\\
&+{\cal O}\big (1/p_1^2\big ).%~~~~~~~~~B9
\label{B9}
\end{align}%{eqnarray}
For the purposes of the renormalization of the TDP (\ref{28}), it is
very important that the leading terms of this asymptotic behavior do
not depend on different chemical potentials, i.e. the quantity
$\sum_{i=1}^4|p_{0i}|$ at $\mu=\nu=\nu_5=0$ has the same asymptotic
(\ref{B9}).
%%%%%%%%%
This conclusion confirms a rather general statement that counterterms do not depend on external parameters. So the coupling constant $G(\Lambda)$ (35) is enough to renormalize the model. 
%%%%%%%%%%%

  Moreover, we would like to emphasize once again that the
asymptotic behavior (\ref{B9}) does not depend on which of the roots
$R_1$, $R_2$ or $R_3$ of the equation (\ref{B6}) is taken as the
quantity $R$ from the relations (\ref{B5}).\vspace{-0.6cm}

\subsection{Consideration of some particular cases}
\vspace{-0.4cm}

Note that in some particular cases it is possible to solve exactly the
third order auxiliary equation (\ref{B6}) and, as a result, to present
the  quasiparticle energies $p_{0i}$ (or the roots $\eta_i$ of the
polynomial (\ref{B3})) in an explicit analytical form.

{\bf\underline{1. The case $\mu=\nu=\nu_5=0$}}.
It is clear from Eq. (\ref{B6}) and Eq. (\ref{B7}) that at $\nu=\nu_5=0$ we
have $A=C=0$, so $R_{1,2}=0$, $R_3=4(M^2+\Delta^2+p_1^2)$. In this
case we have $R=R_3$ and find that $q=s=r^2/4=M^2+\Delta^2+p_1^2$,
$\eta_{1,2}=\sqrt{M^2+\Delta^2+p_1^2}$ and
$\eta_{3,4}=-\sqrt{M^2+\Delta^2+p_1^2}$. If in addition $\mu=0$, then
\begin{align}%{eqnarray}
&\big (|p_{01}|+|p_{02}|+|p_{03}|+|p_{04}|\big )\Big
|_{\mu=\nu=\nu_5=0}\nonumber\\
&=
4\sqrt{M^2+\Delta^2+p_1^2}.%~~~~~~~~~B10
\label{B10}
\end{align}%{eqnarray}
As was noted above, this quantity at $p_1\to\infty$ is expanded in the form  (\ref{B9}).

{\bf\underline{2. The case $\Delta=0$}}. In this particular case the
exact expression for the set of quasiparticle energies $p_{0i}$ was
already presented in Eq. (\ref{26}). Here we would like to demonstrate how
this result is reproduced in the framework of the procedure
(\ref{B3})-(\ref{B41}).

It is easy to see that at $\Delta=0$ there is an evident root $R_1=4\nu^2$ of the polynomial (\ref{B6}). On this basis we can find exact expressions for the other two its roots,
\begin{align}%{eqnarray}
&R_{2,3}=2(M^2+\nu_5^2+p_1^2)\pm
2\sqrt{(M^2+\nu_5^2+p_1^2)^2-4\nu_5^2p_1^2}\nonumber\\
&=(E_1\pm
E_2)^2,
\label{B11}
\end{align}%{eqnarray}
where
\begin{align}
&E_1=\sqrt{M^2+(p_1+\nu_5)^2}, \nonumber\\
&E_2=\sqrt{M^2+(p_1-\nu_5)^2}.
\label{B12}
\end{align}
If $R_1=4\nu^2$ is taken as the quantity $R$ of the relations (\ref{B5}), then, using Eq. (\ref{B5}) in Eq. (\ref{B41}), we obtain directly the expression (\ref{26}) for the set of quasiparticle energies $P_{0i}$ at $k=k'=0$.

If, e.g., $R=R_3\equiv (E_1+E_2)^2$, then, taking into account the evident relation $E_1^2-E_2^2=4p_1\nu_5$, we have from Eq. (\ref{B5})
\begin{align}%{eqnarray}
&r=E_1+E_2,~~~q=E_1E_2-\nu^2+\nu (E_1-E_2),~~~\nonumber\\
&s=E_1E_2-\nu^2-\nu (E_1-E_2),\nonumber\\
&\frac{r^2}{4}-q=\frac{(E_1-E_2-2\nu)^2}{4},~\frac{r^2}{4}-s=\frac{(E_1-E_2+2\nu)^2}{4}.\label{B13}
\end{align}%{eqnarray}
Using these relations in Eq. (\ref{B41}), we receive for the quasiparticle energies $p_{0i}$ the same set as in (\ref{26}) at $k=k'=0$. Thereby we have demonstrated that the set of roots $\eta_i$ (\ref{B41}) does not depend on which of the solutions $R_1$, $R_2$ or $R_3$ of the equation (\ref{B6}) is used as the quantity $R$ in the relations (\ref{B5}).

{\bf\underline{3. The case $M=0$}}. In a similar way it is possible to show that Eq. (\ref{B6}) at $M=0$ has the following three roots:
\begin{eqnarray}
R_1=4\nu_5^2,~~R_{2,3}=({\cal E}_1\pm {\cal E}_2)^2,\label{B14}
\end{eqnarray}
where
\begin{align}
&{\cal E}_1=\sqrt{\Delta^2+(p_1+\nu)^2},\nonumber\\
&{\cal E}_2=\sqrt{\Delta^2+(p_1-\nu)^2}.\label{B15}
\end{align}
On the basis of each of them, using the relations (\ref{B41}) and (\ref{B5}), one can obtain the set of quasiparticle energies (\ref{27}) at $k=k'=0$.

{\bf\underline{4. The case $\nu_5=\nu$}}. In this particular case Eq. (\ref{B6}) has the following three roots:
\begin{eqnarray}
R_1=4\nu^2,~~R_{2,3}=(\widetilde E_1\pm \widetilde E_2)^2,\label{B16}
\end{eqnarray}
where
\begin{align}%{eqnarray}
&\widetilde E_1=\sqrt{M^2+\Delta^2+(p_1+\nu)^2},~~~\nonumber\\
&\widetilde E_2=\sqrt{M^2+\Delta^2+(p_1-\nu)^2}.\label{B17}
\end{align}%{eqnarray}
Taking for simplicity $R=R_1$ in Eq. (\ref{B5}) and using the relations (\ref{B41}), we have in this case for the quasiparticle energies $p_{0i}$ the following set of values:
\begin{align}%{eqnarray}
&\Big\{p_{01},p_{02},p_{03},p_{04}\Big\}\Big
|_{\nu_5=\nu}\nonumber\\
&=\Big\{-\mu-\nu\pm\sqrt{M^2+\Delta^2+(p_1-\nu)^2},\nonumber\\
&-\mu+\nu\pm\sqrt{M^2+\Delta^2+(p_1+\nu)^2}\Big\}.
\end{align}%{eqnarray}

\end{document}